\documentclass[reprint,preprintnumbers,jcp,showpacs,superscriptaddress]{revtex4-1}

\usepackage{palatino}
\usepackage[latin1]{inputenc}
\usepackage{epsf}
\usepackage{amsmath,amssymb}
\usepackage{latexsym}
\usepackage{calc}
\usepackage{color}
\usepackage{epsfig}
\usepackage[pdftex, pdfstartview = {FitH}]{hyperref}

\def\dulR{{\underline{\underline{\bf R}}}}
\def\dulr{{\underline{\underline{\bf r}}}}
\def\bA{{\bf A}}

\begin{document}

\title{Semiclassical analysis of the electron-nuclear coupling in electronic non-adiabatic processes}
\author{Federica Agostini$^a$, Seung Kyu Min$^{a}$ and E. K. U. Gross$^{a}$\\
\vspace{6pt} $^{a}${\em{Max-Planck Institute of Microstructure Physics, Weinberg 2, D-06120 Halle, Germany}}\\
\vspace{6pt} $^{}$}
\date{\today}
\begin{abstract}
In the context of the exact factorization of the electron-nuclear wave function, the coupling between electrons and nuclei beyond the adiabatic regime is encoded (i) in the time-dependent vector and scalar potentials and (ii) in the electron-nuclear coupling operator. The former appear in the Schr\"odinger-like equation that drives the evolution of the nuclear degrees of freedom, whereas the latter is responsible for inducing non-adiabatic effects in the electronic evolution equation. As we have devoted previous studies to the analysis of the vector and scalar potentials, in this paper we focus on the properties of the electron-nuclear coupling operator, with the aim of describing a numerical procedure to approximate it within a semiclassical treatment of the nuclear dynamics. 
\end{abstract}
\maketitle 

\section{Introduction}
Modelling the dynamical coupling of electrons and nuclei beyond the Born-Oppenheimer (BO), or adiabatic, regime is currently among the most challenging problems in the fields of Theoretical Chemistry and Condensed Matter Physics. Within the BO framework, molecular systems are visualized as a set of nuclei moving on a single potential energy surface that represents the effect of the electrons in a given eigenstate. Many interesting phenomena, however, such as vision~\cite{cerulloN2010, ishidaJPCB2012}, charge separation in organic photovoltaic materials~\cite{rozziNC2013, jailaubekovNM2013} or Joule heating in molecular junctions~\cite{todorovJPCMN2007, horsfield_JPhysCM2004}, occur in non-adiabatic conditions. In these situations, solving exactly the time-dependent Schr\"odinger equation (TDSE) for the coupled system of electrons and nuclear is not feasible, as the computational cost scales exponentially with the number of degrees of freedom. However, since the full quantum treatment requires to represent the problem in terms of adiabatic states and transitions among them in regions of strong non-adiabatic coupling, wave packet propagation techniques have been developed, retaining the quantum character of the nuclear dynamics~\cite{cederbaumCPL1990, cederbaumJCP1992, burghardtJCP1999, worthTCA2003, thossCM2004, mayJCP2006, wastermannJCP2013, thossJCP2003, thossJPCC2010, mengJCP2012, brownPCCP, martinezCPL1996, martinezJPC1996, martinezACR2006}. And these techniques are presently the state of the art in quantum dynamics computational methods, proving the benchmark for approximate methods. In fact, for large systems, the dimensionality of the problem does not allow to employ a quantum mechanical description, thus the only feasible approach is to combine a classical description for the nuclei (or ions) with a quantum treatment of a few other essential degrees of freedom, e.g. electrons or protons . In this context, the question of how to model the coupling between the quantum and classical subsystems still remains open, despite the fact that several schemes~\cite{ehrenfest, pechukasPR1969_2, TSH_1990, Thoss_PRL1997, Miller_JCP1997_2, truhlarFD2004, ciccottiJCP2000, kapralJCP2013, bonellaJCP2005, cokerJCP2012, marxPRL2002, martinezACR2006, tavernelliPCCP2011, wyattJCP2001, burghardtJCP2005, prezhdoPRL2001, tannorJCP2012, burghardtJCP2011,thossJCP2013, ananthJCP2013} have been proposed in the literature trying to settle this issue.

In recent work, we have addressed this problem in the context of the exact factorization of the electron-nuclear wave function~\cite{AMG,AMG2}. In such a treatment of quantum dynamics, the solution of the TDSE is written as a single product of a nuclear wave function and an electronic factor, that parametrically depends on the nuclear configuration. Several advantages of this reformulation have been pointed out. First of all, it has been shown~\cite{steps,long_steps} that the nuclear wave function evolves according to a modified TDSE where a time-dependent vector potential and a time-dependent scalar potential represent the effect of the  electrons on the nuclei, beyond the adiabatic regime. When a classical treatment of the nuclear degrees of freedom is introduced, the coupling to the electrons is exactly represented by the force determined from the gradient of the time-dependent potentials~\cite{long_steps,long_steps_mt}. On the other hand, the electronic factor evolves according to a (less standard) evolution equation, coupled to the nuclear TDSE, where the effect of the nuclei is represented by an \textsl{electron-nuclear coupling operator} explicitly depending on the nuclear wave function.

The analysis of the time-dependent potentials and of the classical nuclear force has been the subject of previous work. We have been able to analyze a simple model system to pinpoint some relevant features~\cite{steps,long_steps,long_steps_mt} of the potentials that should be accounted for when developing approximations. Based on these observations, starting from the exact formulation, we have proposed~\cite{mqc, long_mqc} a novel mixed quantum-classical algorithm to solve the coupled electronic and nuclear evolution equations in a fully approximated way. The classical limit is considered as the lowest-order, in a $\hbar$-expansion, of the nuclear wave function in the complex-phase representation~\cite{VanVleck_PNAS1928}.

In the present paper, the focus is directed towards the analysis of the \textsl{electron-nuclear coupling operator} that, in the electronic equation, mediates the coupling to the nuclei and that depends explicitly on (the gradient of) the nuclear wave function. It is fundamental to be able to correctly approximate such term, as it is responsible to induce electronic non-adiabatic transitions~\cite{mqc,long_mqc} and decoherence~\cite{MQC_min}. When a classical description of the nuclei is adopted, the concept of wave function is somehow lost and problems arise when approximating this operator. If a distribution of trajectories~\cite{long_steps_mt} is used to mimic the evolution of the nuclear wave function, its modulus and phase cannot be smooth functions of space. The numerical error thus introduced affects the calculations, but it can be cured if refined approximations are considered. The goal of this paper is to describe a procedure to avoid the above issue and to test its efficiency against exact calculations for a simple model system. Therefore, we propose here (i) to employ a representation of the nuclear density in terms of evolving frozen gaussians (FGs)~\cite{Heller_JCP1981_2}, rather than trajectories, following the scheme presented in Ref.~\cite{long_steps_mt} and (ii) to estimate the phase of the nuclear wave function adopting such a FGs picture, based on a simplified form of the semiclassical Herman-Kluk~\cite{Kluk_CP1984, Davis_JCP1986, Kay_CP2006, Miller_MP2002} propagation scheme within the initial value representation (IVR) theory~\cite{Miller_JCP1970, Wang_ARPC2004, Kay_ARPC2005, Miller_JPCA2001}. The model system for non-adiabatic charge transfer of Ref.~\cite{MM} allows for an exact numerical solution of the full quantum mechanical problem, thus providing a benchmark to any approximation that will be considered. Starting from these results, we will compute the exact time-dependent scalar potential, also referred to as time-dependent potential energy surface (TDPES), in a gauge where the vector potential can be set to zero. The effect of the electrons is then fully accounted for by this TDPES, that is adopted to evolve the FGs. Since the electronic part of the problem is solved exactly, the only source of error will be in this semiclassical approximation. It is worth stressing that this procedure does not result in the development of a new algorithm, but is a test of the performance of the FGs approximation of the nuclear motion.

The paper is organized as follows. In Section~\ref{sec: factorization} we briefly recall the factorization formalism and we focus on the analysis of the electron-nuclear coupling (ENC) operator in the electronic evolution equation. Section~\ref{sec: semiclassics} is devoted to a discussion on the apronximations employed in the calculations. Numerical results are presented in Section~\ref{sec: results}, by comparing situations with different non-adiabatic coupling strengths and testing the approximations employed to evaluate the ENC term. Conclusions are stated in Section~\ref{sec: conclusion}.

\section{The exact factorization framework}\label{sec: factorization}
In the absence of an external field, the non-relativistic Hamiltonian $ \hat H = \hat T_n+\hat H_{BO}$ describes a system of interacting nuclei and electrons. Here, $\hat T_n$ denotes the nuclear kinetic energy and $\hat H_{BO}(\dulr,\dulR)=\hat 
T_e(\dulr)+\hat V_{e,n}(\dulr,\dulR)$ is the BO Hamiltonian, containing the electronic kinetic energy $\hat T_e(\dulr)$ and all interactions $\hat V_{e,n}(\dulr,\dulR)$. As recently proven~\cite{AMG, AMG2}, the full wave function, $\Psi(\dulr,\dulR,t)$, solution of the TDSE
\begin{equation}\label{eqn: tdse}
 \hat H\Psi(\dulr,\dulR,t)=i\hbar\partial_t\Psi(\dulr,\dulR,t),
\end{equation}
can be written as the product 
\begin{equation}\label{eqn: factorization}
 \Psi(\dulr,\dulR,t)=\Phi_\dulR(\dulr,t)\chi(\dulR,t),
\end{equation}
of the nuclear wave function, $\chi(\dulR,t)$, and the electronic wave function, $\Phi_\dulR(\dulr,t)$, which parametrically depends of the nuclear 
configuration~\cite{hunter, Gross_PTRSA2014}. Throughout the paper the symbols $\dulr,\dulR$ indicate the coordinates of the $N_e$ electrons and 
$N_n$ nuclei, respectively. Eq.~(\ref{eqn: factorization}) is unique under the partial normalization condition (PNC)
\begin{equation}\label{eqn: pnc}
 \int d\dulr\left|\Phi_\dulR(\dulr,t)\right|^2=1\,\,\,\forall \,\dulR,t
\end{equation}
up to within a gauge-like phase transformation. The evolution equations for $\Phi_\dulR(\dulr,t)$ and $\chi(\dulR,t)$,
\begin{align}
 \left(\hat H_{el}-\epsilon(\dulR,t)\right)\Phi_\dulR(\dulr,t) = i\hbar \partial_t\Phi_\dulR(\dulr,t)\label{eqn: electronic eqn} \\
 \hat H_n\chi(\dulR,t) = i\hbar \partial_t\chi(\dulR,t),\label{eqn: nuclear eqn}
\end{align}
are derived by applying Frenkel's action principle~\cite{frenkel,mclachlan} with respect to the two wave functions and are exactly equivalent~\cite{AMG, AMG2} to the TDSE~(\ref{eqn: tdse}). Eqs.~(\ref{eqn: electronic eqn}) and~(\ref{eqn: nuclear eqn}) are obtained by imposing the PNC~\cite{alonsoJCP2013, AMG2013} by means of Lagrange multipliers.

The electronic equation~(\ref{eqn: electronic eqn}) contains the electronic Hamiltonian
\begin{equation}
\hat H_{el}=\hat H_{BO}(\dulr,\dulR) + \hat U_{en}^{coup}[\Phi_\dulR,\chi],
\end{equation}
which is the sum of the BO Hamiltonian and the ENC operator $\hat U_{en}^{coup}[\Phi_\dulR,\chi]$,
\begin{align}
\hat U_{en}^{coup}&[\Phi_\dulR,\chi]=\sum_{\nu=1}^{N_n}\frac{1}{M_\nu}\left[
 \frac{\left[-i\hbar\nabla_\nu-\bA_\nu(\dulR,t)\right]^2}{2} \right.\label{eqn: enco} \\
& \left.+\left(\frac{-i\hbar\nabla_\nu\chi}{\chi}+\bA_\nu(\dulR,t)\right)
 \left(-i\hbar\nabla_\nu-\bA_{\nu}(\dulR,t)\right)\right].\nonumber
\end{align}
In Eq.~(\ref{eqn: electronic eqn}), $\epsilon(\dulR,t)$ is the TDPES, defined as
\begin{equation}\label{eqn: tdpes I}
\epsilon(\dulR,t)=\left\langle \Phi_\dulR(t)\right|\hat H_{el}-i\hbar\partial_t\left|\Phi_\dulR(t)\right\rangle_\dulr.
\end{equation}
$\hat U_{en}^{coup}$ and $\epsilon(\dulR,t)$, along with the vector potential $\bA(\dulR,t)$,
\begin{equation}\label{eqn: vector potential}
\bA(\dulR,t) = \left\langle \Phi_\dulR(t)\right| \left. -i\hbar\nabla_\nu\Phi_\dulR(t)\right\rangle_\dulr,
\end{equation}
mediate the coupling between electrons and nuclei in a formally exact way. Here, the symbol $\langle\,\cdot\,|\,\cdot\,\rangle_\dulr$ stands for an integration over electronic coordinates.

The nuclear evolution is generated by the Hamiltonian
\begin{equation}\label{eqn: nuclear H}
\hat H_n(\dulR,t) = \sum_{\nu=1}^{N_n} \frac{\left[-i\hbar\nabla_\nu+\bA_\nu(\dulR,t)\right]^2}{2M_\nu} + \epsilon(\dulR,t),
\end{equation}
according to the TDSE~(\ref{eqn: nuclear eqn}).

The TDPES and the vector potential are uniquely determined up to within gauge-like transformations~\cite{AMG,AMG2}. The uniqueness can be straightforwardly proven by following the steps of the current density version~\cite{Ghosh-Dhara} of the  Runge-Gross theorem~\cite{RGT}. In this paper, as a choice of gauge, we introduce the additional constraint ${\bf A}_{\nu}\left(\dulR,t\right)=0$ (see Ref.~\cite{long_steps_mt} for a detailed discussion on how this condition can be imposed)~\cite{CI_MAG}.

As discussed in the introduction, we study the properties of the ENC term $-i\hbar\nabla_\nu\chi/\chi$ that in the expression of the operator $\hat U_{en}^{coup}[\Phi_\dulR,\chi]$ explicitly depends on the nuclear wave function. This analysis is based on the interest in developing a procedure to approximate it when a classical or semiclassical treatment of the nuclear motion is adopted. For instance, in the classical limit, we have derived~\cite{mqc,long_mqc} its expression in terms of the nuclear momentum. This has been done by writing the nuclear wave function as $\chi(\dulR,t)=\exp{\left[\frac{i}{\hbar}\mathcal S(\dulR,t)\right]}$, with $\mathcal S(\dulR,t)$ a complex function~\cite{VanVleck_PNAS1928}. If we now suppose~\cite{mqc, long_mqc} that this function can be expanded as an asymptotic series in powers of $\hbar$, namely $\mathcal S(\dulR,t)=\sum_{\alpha}\hbar^{\alpha}S_{\alpha}(\dulR,t)$, the ENC term becomes
\begin{equation}\label{eqn: zeroth order of nabla chi / chi}
\frac{-i\hbar\nabla_\nu\chi(\dulR,t)}{\chi(\dulR,t)} = \nabla_\nu S_0(\dulR,t),
\end{equation}
at the lowest-order in $\hbar$. On the right-hand-side (RHS), the function $\nabla_\nu S_0(\dulR,t)$ is the classical nuclear momentum evaluated along the trajectory, since~\cite{mqc,long_mqc} $S_0$ satisfies a Hamilton-Jacobi equation with Hamiltonian
\begin{equation}\label{eqn: classical hamiltonian}
 H_n = \sum_{\nu=1}^{N_n}\frac{\left[\nabla_\nu S_0(\dulR,t)+{\bf A}_\nu(\dulR,t)\right]^2}{2M_\nu}+ \epsilon(\dulR,t).
\end{equation}
The vector potential appears in the above expression of the classical Hamiltonian because this result has general validity, not only in the gauge adopted in the following calculations. 

Alternatively, if the nuclear wave function is written in terms of its modulus and phase, $\chi=|\chi|e^{iS/\hbar}$, the (exact) expression of the ENC term becomes
\begin{align}\label{eqn: polar nabla chi over chi}
 \frac{-i\hbar\nabla_\nu\chi(\dulR,t)}{\chi(\dulR,t)} = \nabla_\nu S(\dulR,t) +i\frac{-\hbar\nabla_\nu\left|\chi(\dulR,t)\right|}
 {\left|\chi(\dulR,t)\right|}.
\end{align}
It is clear at this point that a good estimate of the ENC term is only possible when both the modulus and the phase of the nuclear wave function are correctly described. A classical treatment, as in Eq.~(\ref{eqn: zeroth order of nabla chi / chi}), only provides an approximation to the real part of the ENC term, while the information about the imaginary part is lost~\footnote{To be precise, the nuclear density is approximated, within the classical treatment, as $\delta$-function, centered at all times at the classical trajectory. But this contribution is totally omitted in the expression of the ENC term.}.

In the following, we will introduce a FG-based approach to determine an approximation to Eq.~(\ref{eqn: polar nabla chi over chi}). FGs~\cite{Heller_JCP1981_2} evolving on the exact TDPES are used to reconstruct the nuclear density, thus allowing to calculate the second term on the RHS, as $|\chi|$ is a smooth function of the nuclear coordinates. The phase information is instead encoded in the classical action accumulated over time and associated to each FG, as we will show in Section~\ref{sec: semiclassics}.

Henceforth, we will drop the bold-double underlined notation for electronic and nuclear positions as we will deal with one-dimensional (1D) quantities.

\section{Semiclassical approximation}\label{sec: semiclassics}
\subsection{Nuclear density}\label{sec: nuclear density}
According to the procedure presented in Ref.~\cite{long_steps_mt}, a set of independent classical trajectories evolving on the exact TDPES are able to reproduce the nuclear density in almost perfect agreement with quantum results. As pointed out in the Introduction, however, constructing a histogram from the distribution of the trajectories does not allow to compute the second term on the RHS of Eq.~(\ref{eqn: polar nabla chi over chi}), that involves the gradient of $|\chi|$, without a large numerical error. The reason is that the ``classical'' density is not a smooth function of space. The solution proposed here is to improve the previous approximation of nuclear dynamics, by propagating the mean positions and momenta of a set of FGs on the exact TDPES, rather than classical trajectories.

Given a set of $N_{traj}$ initial positions and momenta, $R_0,P_0$, sampled as described in Section~\ref{sec: results}, complex gaussians, also referred to as \textsl{coherent states}, are constructed as
\begin{align}\label{eqn: coherent states}
 g\left(R;R_{l,0}P_{l,0},\gamma\right) =
 \left(\frac{\gamma}{\pi}\right)^{\frac{1}{4}}e^{-\frac{\gamma}{2}\left(R-R_{l,0}\right)^2+\frac{i}{\hbar}P_{l,0}(R-R_{l,0})},
\end{align}
with width $\gamma$ to be determined below. Each FG is also associated a ``weight'',
\begin{align}\label{eqn: weight}
w_l = \int dRg^*\left(R;R_{l,0}P_{l,0},\gamma\right)\chi_0(R),
\end{align}
corresponding to the projection of the initial nuclear wave function $\chi_0(R)$ on the FGs. The nuclear density at each time is then obtained as
\begin{align}\label{eqn: density as sum of gaussians}
\left|\chi(R,t)\right|^2 \simeq \sum_{l=1}^{N_{traj}} |w_l|^2\left|g\left(R;R_{l}(t)P_{l}(t),\gamma\right)\right|^2,
\end{align}
where $R_l(t),P_l(t)$ are the time-evolved positions and momenta of $R_{l,0},P_{l,0}$. In comparison to the purely classical approximation, Eq.~(\ref{eqn: density as sum of gaussians}) allows not only to reproduce the nuclear density in very good agreement with quantum results, as will be shown in Section~\ref{sec: results}, but also to calculate (analytically) the gradient of the nuclear density (or of the modulus, as it appears in Eq.~(\ref{eqn: polar nabla chi over chi})). 

It is important to notice that Eq.~(\ref{eqn: density as sum of gaussians}) is an approximation to the nuclear density when the nuclear wave function is represented as a superposition of coherent states. In fact, coherent states form an overcomplete basis and, in writing Eq.~(\ref{eqn: density as sum of gaussians}), we neglect the overlaps of coherent states. The reason for this further approximation is related to the choice of the initial set of positions and momenta, the mean positions and mean momenta of the FGs. On one hand, we want to maintain the same choice of initial conditions done for the classical propagation (see Section~\ref{sec: results}), in order to be able to directly compare classical and FG results. On the other hand, we want to obtain an initial nuclear density as close as possible to the exact density. We have thus computed the root mean square deviation (RMSD) for the nuclear density when either (1) Eq.~(\ref{eqn: density as sum of gaussians}) is employed or (2) the full expression is considered, i.e. with overlap terms. The best agreement achieved in case (1) is for $\gamma=$ 7.0~a$_0^{-2}$ (used in Section~\ref{sec: results}), with RMSD = 0.020, 0.017, 0.006 for $N_{traj}=$ 2000, 5000, 10000, respectively, whereas in case (2) is for $\gamma=$ 3.0~a$_0^{-2}$ with RMSD = 0.037, 0.024 0.0013. It is evident that a better agreement at the initial time is obtained for case (1), namely when the approximation in Eq.~(\ref{eqn: density as sum of gaussians}) is used, for all the values of $N_{traj}$. Once the parameters defining the coherent states are selected, namely $R_l, P_l,\gamma$, the weights $w_l$ associated to each FG are automatically determined by Eq.~(\ref{eqn: weight}) and kept constant throughout the propagation.

\subsection{Nuclear phase}
The phase of the nuclear wave function will be determined according to 
\begin{align}\label{eqn: semiclassical phase}
S(R,t)\simeq\arctan\frac{\Im \left[\chi_{SC}(R,t)\right]}{\Re \left[\chi_{SC}(R,t)\right]},
\end{align}
where $\chi_{SC}(R,t)$ is a semiclassical approximation to the exact $\chi(R,t)$. Following the Herman-Kluk procedure~\cite{Kluk_CP1984, Davis_JCP1986, Kay_CP2006, Miller_MP2002} to approximate the quantum propagator, the expression of the time-evolved wave function at time $t$ is
\begin{align}\label{eqn: semiclassical propagator}
\chi_{SC}(R,t)= \int \frac{dR_0dP_0}{2\pi\hbar}C_t&(R_0,P_0)  e^{\frac{i}{\hbar}S(R_0,P_0;t)}\nonumber \\
 &\left\langle R|R_tP_t,\gamma\right\rangle\left\langle R_0P_0,\gamma|\chi_0\right\rangle.
\end{align}
Here, $\left\langle R_0P_0,\gamma|\chi_0\right\rangle$ denotes the projection of the initial nuclear wave function on the coherent states, similarly to Eq.~(\ref{eqn: weight}), while $\left\langle R|R_tP_t,\gamma\right\rangle$ is an alternative expression for the coherent states (see Eq.~(\ref{eqn: coherent states})). $R_t,P_t$ are the (classically) evolved positions and momenta corresponding to the initial conditions $R_0,P_0$. $S(R_0,P_0;t)$ is the classical action accumulated up to time $t$ along the trajectory whose initial conditions are $R_0,P_0$. The Herman-Kluk pre-factor is indicated here with the symbol $C_t(R_0,P_0)$~\cite{Kluk_CP1984, Davis_JCP1986, Kay_CP2006, Miller_MP2002}, but it will be set equal to unity throughout the calculations. Therefore, the symbol $\chi_{SC}(R,t)$ will be replaced by $\chi_{FG}(R,t)$, since we will use a FGs approximation rather than a rigorous semiclassical approximation.

In the procedure employed here, the semiclassical nuclear wave function is estimated as a sum over trajectories of time-evolved coherent states, namely 
\begin{align}
\chi_{FG}(R,t) = \sum_{l=1}^{N_{traj}} w_l\frac{e^{\frac{i}{\hbar}S_l}}{2\pi\hbar}g(R;R_l(t),P_l(t),\gamma),
\end{align}
where the $l$-th classical action is calculated as
\begin{align}
 S_l = \int_0^t d\tau \left(\frac{P_l^2(\tau)}{2M}-\epsilon(R_l(\tau),\tau)\right).
\end{align}
Here, the (exact) TDPES is evaluated, at time $\tau$, at the classical position $R_l(\tau)$ and its expression is obtained by solving the TDSE for the full wave function $\Psi$ and by directly calculating Eq.~(\ref{eqn: tdpes I}) once the factorization~(\ref{eqn: factorization}) is applied.

\section{Numerical results}\label{sec: results}
\begin{figure}[h!]
 \centering
 \includegraphics*[width=.35\textwidth]{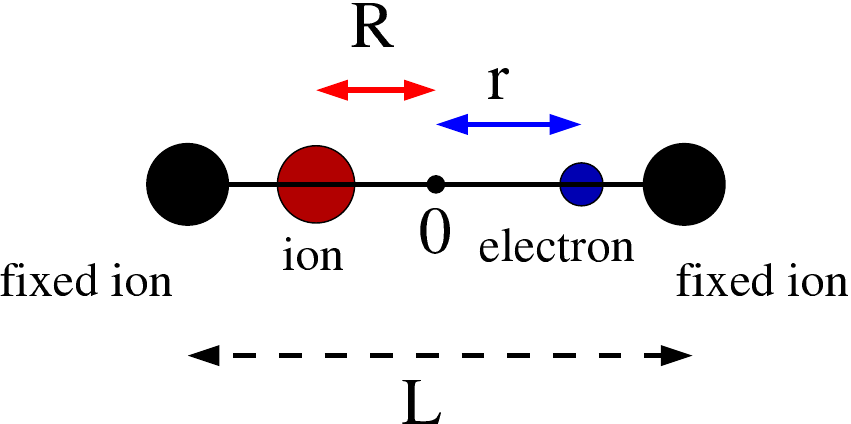}
 \caption{Schematic representation of the model system described by the Hamiltonian~(\ref{eqn: metiu-hamiltonian}).}
 \label{fig: metiu model}
\end{figure}

\begin{figure}[h!]
 \begin{center}
 \includegraphics[angle=270,width=.4\textwidth]{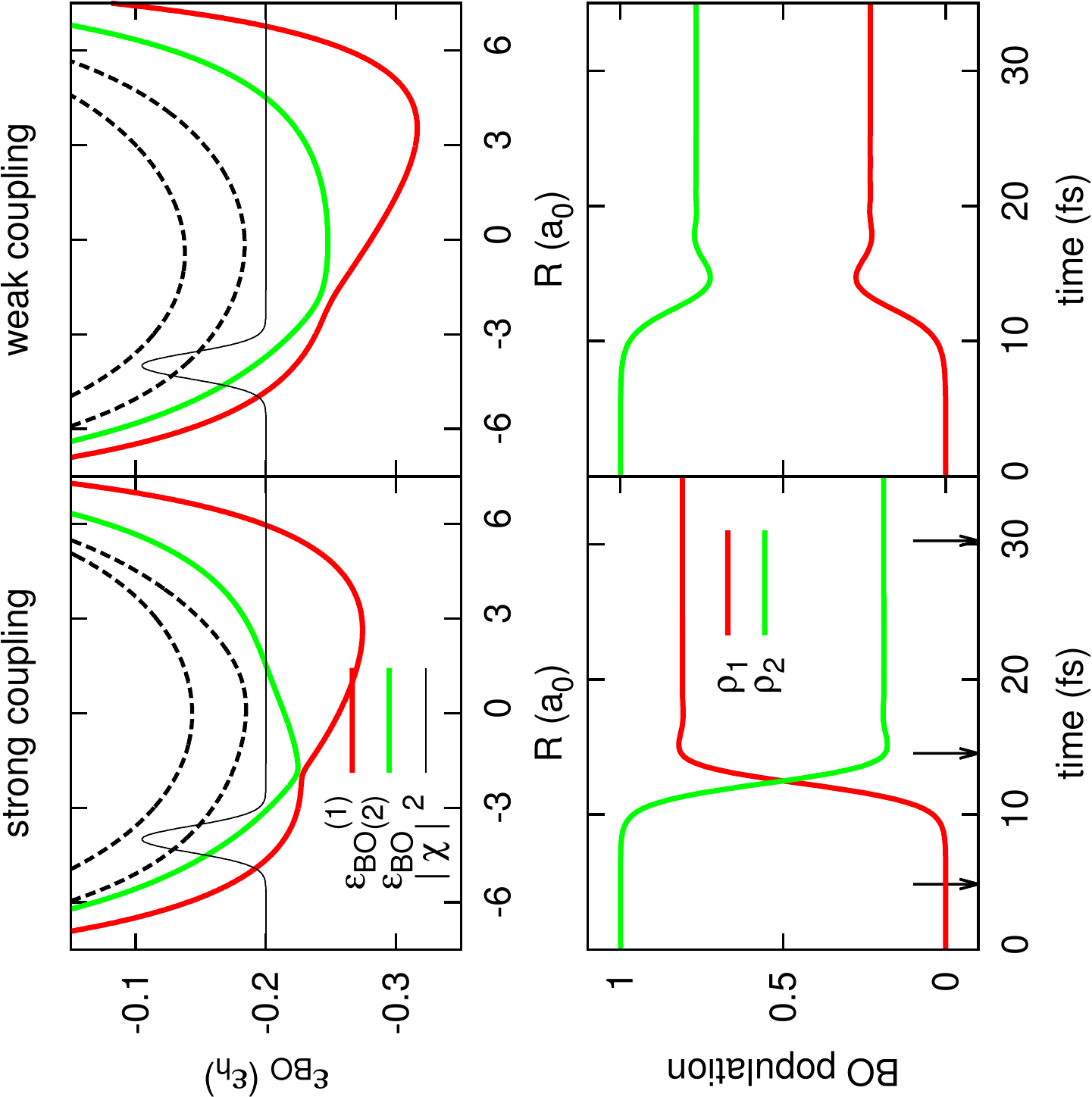}
 \caption{Upper panels: Lowest four BO surfaces, as functions of the nuclear coordinate, for strong (left) and weak (right) non-adiabatic coupling strengths. The first (red line) and second (green line) surfaces correspond to the adiabatic states that are populated during the dynamics, whereas the third and fourth surfaces (dashed black lines) are shown for reference. The squared modulus (reduced by ten times and rigidly shifted in order to superimpose it on the energy curves) of the initial nuclear wave packet is also shown (thin black line). Lower panels: Populations of the two lowest adiabatic states ($\rho_1$ and $\rho_2$) as functions of time. The arrows represent the time-steps shown in the following figures.}
 \label{fig: BO}
 \end{center}
\end{figure}

The expression of the TDPES according to Eq.~(\ref{eqn: tdpes I}) is determined by calculating the electronic wave function $\Phi_R(r,t)$ from the 
full wave function $\Psi(r,R,t)$, which is known at all times by solving the TDSE~(\ref{eqn: tdse}) for the model Hamiltonian
\begin{align}
 \hat{H}(r,R)= -\frac{1}{2}\frac{\partial^2}{\partial r^2}-\frac{1}{2M}\frac{\partial^2}{\partial R^2}  +
 \frac{1}{\left|\frac{L}{2}-R\right|}+\frac{1}{\left|\frac{L}{2} + R\right|}\label{eqn: metiu-hamiltonian}\\
 -\frac{\mathrm{erf}\left(\frac{\left|R-r\right|}{R_f}\right)}{\left|R - r\right|}
 -\frac{\mathrm{erf}\left(\frac{\left|r-\frac{L}{2}\right|}{R_r}\right)}{\left|r-\frac{L}{2}\right|}
 -\frac{\mathrm{erf}\left(\frac{\left|r+\frac{L}{2}\right|}{R_l}\right)}{\left|r+\frac{L}{2}\right|}.\nonumber
\end{align}
This system has been introduce by Shin and Metiu~\cite{MM} as a prototype for non-adiabatic charge transfer. The system is 1D and consists of three 
ions and a single electron, as depicted in Fig.~\ref{fig: metiu model}.
Two ions are fixed at a distance of $L=19.0$~$a_0$, the third ion and the electron are free to move in 1D along the line joining the two fixed ions. Here, the symbols $r$ and $R$ are the coordinates of the electron and the movable ion measured from the center of the two fixed ions. The ionic mass is chosen as $M=1836$, the proton mass, whereas the other parameters are tuned in order to make the system essentially a two-electronic-state model. We present here the results obtained by choosing two different sets of parameters, producing strong and weak non-adiabatic couplings, between the first, $\epsilon^{(1)}_{BO}$, and the second BOPES, $\epsilon^{(2)}_{BO}$, around the avoided crossing at $R_{ac}=-1.90$~a$_0$. The values of the parameters in the Hamiltonian~(\ref{eqn: metiu-hamiltonian}) are: $R_f=5.0$~a$_0$, $R_l=3.1$~a$_0$ and $R_r=4.0$~a$_0$, for the strong coupling case; $R_f=3.8$~a$_0$, $R_l=2.9$~a$_0$ and $R_r=5.5$~a$_0$, for the weak coupling case. The BO surfaces are shown in Fig.~\ref{fig: BO} (upper panels).

We study the time evolution of this system by choosing the initial wave function as the product of a real-valued normalized Gaussian wave packet, centered at $R_c=-4.0$~$a_0$ with variance $\sigma=1/\sqrt{2.85}$~$a_0$ (thin black line in Fig.~\ref{fig: BO}, upper panels), and the second BO electronic state. To calculate the TDPES, we first solve the TDSE~(\ref{eqn: tdse}) for the complete system, with Hamiltonian~(\ref{eqn: metiu-hamiltonian}), and obtain the full wave function, $\Psi(r,R,t)$. This is done by numerical integration of the TDSE using the split-operator-technique~\cite{spo}, with time-step of $2.4\times10^{-3}$~fs (or $0.1$~a.u.).

The mean positions and momenta of the FGs evolve along classical trajectories, generated according to Hamilton's equations
\begin{align}
 \dot R_l(t) = P_l(t)/M;\,\,\dot P_l(t) = -\partial_R \epsilon\left(R,t\right)\big|_{R_l(t)},
  \label{eqn: rp dot}
\end{align}
integrated by using the velocity-Verlet algorithm with a time-step of $12.0\times10^{-3}$~fs (or $0.5$~a.u.). The initial conditions are sampled from the Wigner phase-space distribution corresponding to the initial nuclear wave function. The initial coherent states used in Eq.~(\ref{eqn: weight}) are thus constructed to determine the (complex-valued) weights $w_l$. Numerical results have been obtained for sets of $N_{traj}=2000, 5000, 10000$ trajectories, with the value~\footnote{The value of $\gamma$ is chosen to have the best overlap between the initial semiclassical density and the exact density, in order to impose initial conditions as close as possible to the exact ones.} $\gamma=7.0$~a$_0^{-2}$ for the width of the FGs. The coherent states used to represent the nuclear wave function form an overcomplete basis, therefore the sum in Eq.~(\ref{eqn: density as sum of gaussians}) has to be truncated. In order to choose the adequate number of basis functions to include in the sum, energy conservation has been tested and confirmed for all values of $N_{traj}$. The results will be presented only for the case $N_{traj}=5000$. We have computed the RMSD between the exact nuclear density and its approximation in Eq.~(\ref{eqn: density as sum of gaussians}) for a set of values of $\gamma$ and we have chosen the value of this parameter for which the RMSD is minimum (see also the discussion at the end of Section~\ref{sec: nuclear density}), but clearly other values can be selected.

We will confirm below that the semiclassical FG scheme, adapted to the factorization approach, is an accurate and efficient way to approximate the nuclear wave function. Before presenting the FG results, let us first show different snapshots taken along the dynamics, showing the TDPES. Its feature have been extensively discussed in previous work~\cite{steps,long_steps,long_steps_mt}, but for the sake of completeness, we report here a few configurations. Moreover, we would like to underline that here results for different non-adiabatic coupling strengths will be presented, in order to test the 
efficiency of the method also in the weak coupling regime.
\begin{figure}[h!]
 \begin{center}
 \includegraphics[angle=270,width=.4\textwidth]{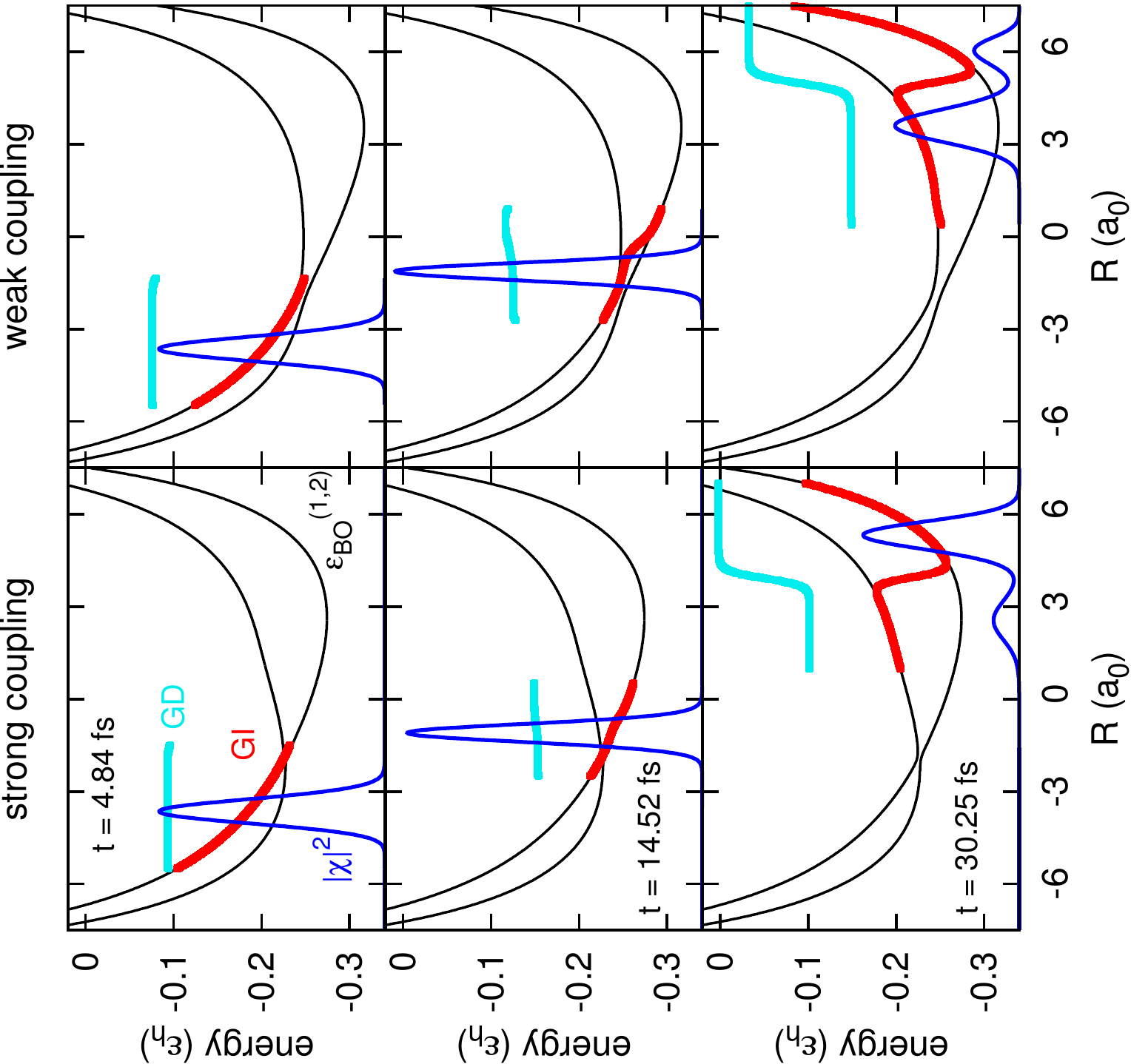}
 \caption{TDPES at different time-steps, as indicated by the arrows in Fig.~\ref{fig: BO}, for strong (left) and weak (right) non-adiabatic coupling strengths. The two lowest BO surfaces (black) are plotted for reference. The GI part of the TDPES (red) presents the well-studied~\cite{steps,long_steps} dynamical steps that bridge piecewise adiabatic shapes potential energy surface, whereas the GI part~\cite{long_steps_mt} (cyan) is piecewise constant. The nuclear density (blue) is shown for reference.}
 \label{fig: tdpes}
 \end{center}
\end{figure}

Fig.~\ref{fig: tdpes} shows the gauge-invariant (GI) and gauge-dependent (GD) components of the TDPES, namely the two terms that can be identified in Eq.~(\ref{eqn: tdpes I}) as $\epsilon_{GI}\left(\dulR,t\right)=\langle \Phi_\dulR(t)|\hat H_{el}|\Phi_\dulR(t)\rangle_\dulr$ and $\epsilon_{GD}\left(\dulR,t\right)=\langle \Phi_\dulR(t)|-i\hbar\partial_t|\Phi_\dulR(t)\rangle_\dulr$. The snapshots shown in Fig.~\ref{fig: tdpes} are taken along the evolution at the times indicated by the arrows in Fig.~\ref{fig: BO} (lower panels) and the two lowest adiabatic surfaces are shown for reference. As previously discussed~\cite{steps,long_steps,long_steps_mt}, before the splitting of the nuclear wave packet, the GI part of the TDPES is diabatic, whereas the GD part is constant, and after the splitting the GI component develops steps that bridge between different adiabatic surfaces, whereas the GD part is piecewise constant. The TDPES is only calculated in the regions where the nuclear density is (numerically) not zero. The lack of reliable information beyond the regions shown in the figures is not an issue when classical trajectories or FGs are employed to mimic the nuclear density, as the regions where the exact density is exponentially small are not, or are poorly, sampled.

It is evident from Fig.~\ref{fig: tdpes} that we will discuss results for short dynamics, limited to the first half of the oscillation period of the nuclear wave packet in the potential well. Interesting dynamics may arise at later times, when for instance the nuclear wave packet crosses a second time the non-adiabatic coupling region. However, here we focus on the initial non-adiabatic event and test how the FG approximation capture this process. Due to the fact that the analysis reported below is the first attempt to incorporate semiclassically nuclear quantum effects in the exact factorization formalism, we study a simple situation that nonetheless captures the main features of a non-adiabatic event. Also, situations where, for instance, reproducing tunneling dynamics might represent a problem for classically evolving FGs~\cite{:/content/aip/journal/jcp/129/2/10.1063/1.2949818,ankerhold_book,AnkerholdPRA2003,KayPRA2013} will be the subject of further study and not addressed here.

\begin{figure}[h!]
\begin{center}
\includegraphics[angle=270,width=.4\textwidth]{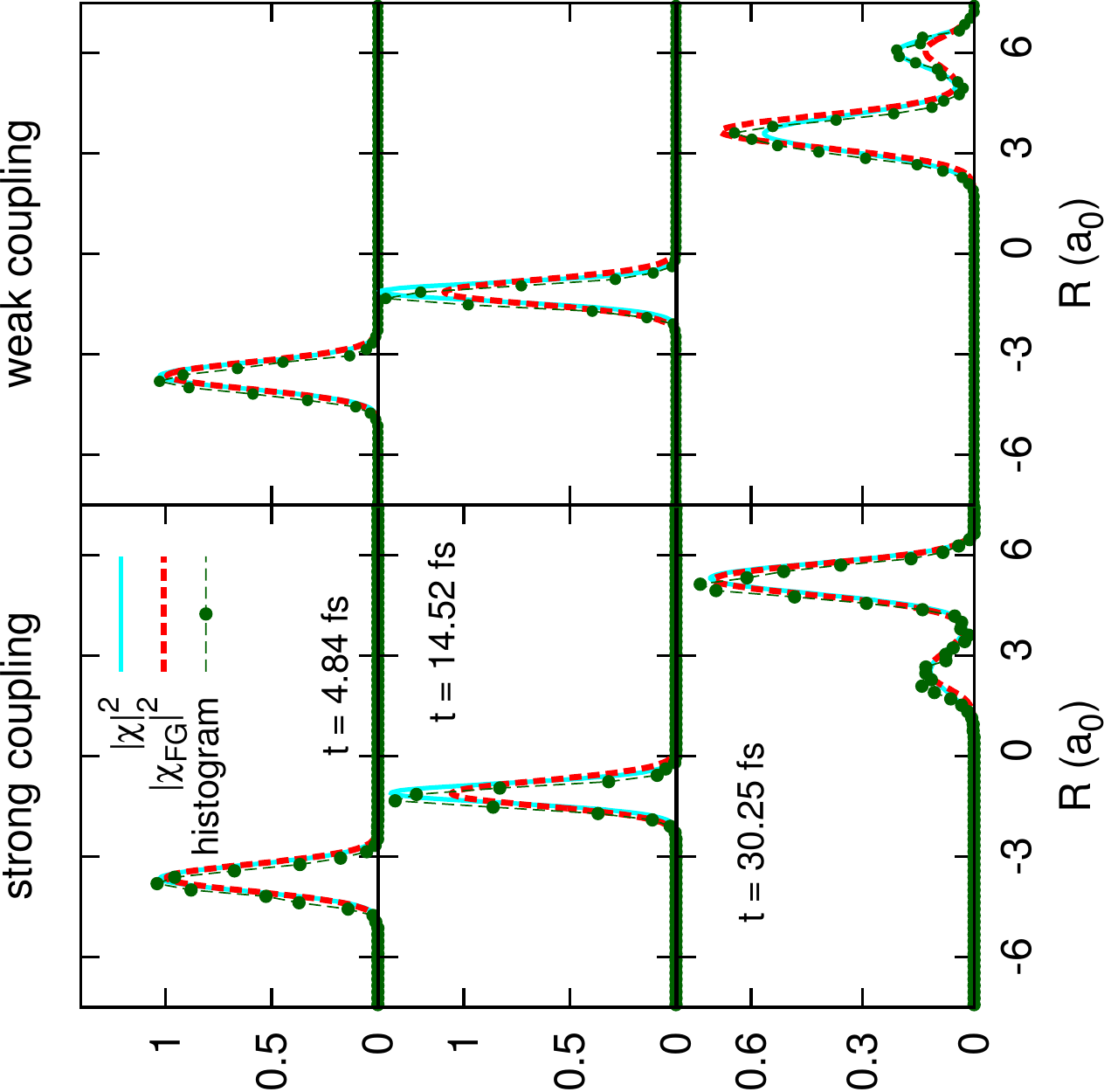}
\caption{Nuclear density for strong (left) and weak (right) non-adiabatic coupling strengths. Exact results (cyan) are compared with the 
semiclassical density (red), expressed as sum of FGs, and with the histogram (green) constructed from the distribution of classical positions 
along the trajectories.}
\label{fig: nwf2}
\end{center}
\end{figure}

A first set of results is shown in Fig.~\ref{fig: nwf2}, where we compare the nuclear density, $|\chi(R,t)|^2$, from three different calculations: quantum (cyan), employing the full electron-nuclear wave function; classical (green), where the histogram is constructed from the distribution of classical trajectories evolving on the TDPES according to Eqs.~(\ref{eqn: rp dot}) (as in Ref.~\cite{long_steps_mt}); FG (red), with the nuclear density given in Eq.~(\ref{eqn: density as sum of gaussians}). 
As expected from previous calculations~\cite{long_steps_mt}, the use of classical trajectories seems to be enough accurate to reproduce the nuclear density. It is however important to stress again, that these results are not obtained by solving a fully approximate form of the coupled electronic and nuclear equations~(\ref{eqn: electronic eqn}) and~(\ref{eqn: nuclear eqn}). They only represent a benchmark for any quantum-classical algorithm, since the effect of the electrons, via the TDPES, is treated exactly. The semiclassical density, constructed as the weighted sum of FGs given in Eq.~(\ref{eqn: density as sum of gaussians}), is also accurate. In comparison to the classical histogram, the gain here is the smoothness of the density, not achievable with purely classical trajectories. This feature is extremely important for the calculation of the ENC term containing the gradient of the nuclear density, via the term $-i\hbar\nabla_\nu|\chi|/|\chi|$ in Eq.~(\ref{eqn: polar nabla chi over chi}).

\begin{figure}[h!]
\begin{center}
\includegraphics[angle=270,width=.4\textwidth]{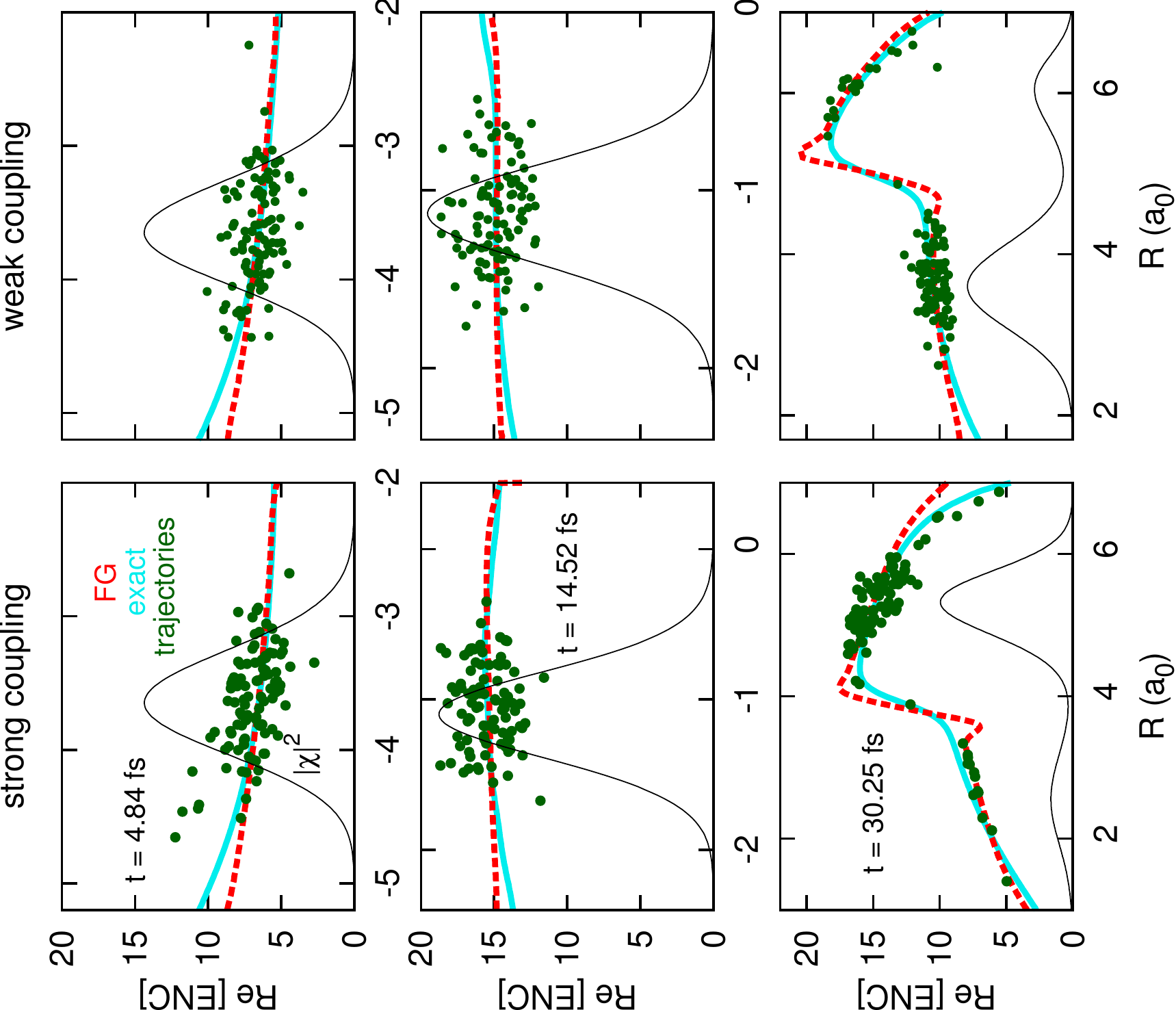}
\caption{Real part of the ENC term for strong (left) and weak (right) non-adiabatic coupling strengths. The nuclear density (black) is shown for reference. Exact results (cyan) are compared with semiclassical calculations (red) and with the classical phase-space points (green).}
\label{fig: Re ENC}
\end{center}
\end{figure}

The second important characteristic of the semiclassical approach is that each FG contributes a phase factor to the full nuclear wave function, thus allowing to determine also the first term on the RHS of Eq.~(\ref{eqn: polar nabla chi over chi}). We show this term, i.e. $\nabla_\nu S(R,t)$, in Fig.~\ref{fig: Re ENC}, comparing once again exact results with the corresponding FG and classical approximations. The semiclassical value of $\nabla_\nu S(R,t)$ is determined as the gradient of the phase in Eq.~(\ref{eqn: semiclassical phase}), whereas the function $\nabla_\nu S(R,t)$ is classically interpreted as the nuclear momentum evaluated along each trajectory. While the agreement between quantum and FG results is remarkable, the phase-space points corresponding to the classical trajectories do not allow to reconstruct a smooth function of $R$. Even if the number of classical trajectories is increased, the phase-space points are too ``noisy'' to allow for reconstructing a smooth function. A smoothing algorithm should then be employed, but the numerical efficiency of the whole procedure might become questionable.

\begin{figure}[t!]
\begin{center}
\includegraphics[angle=270,width=.4\textwidth]{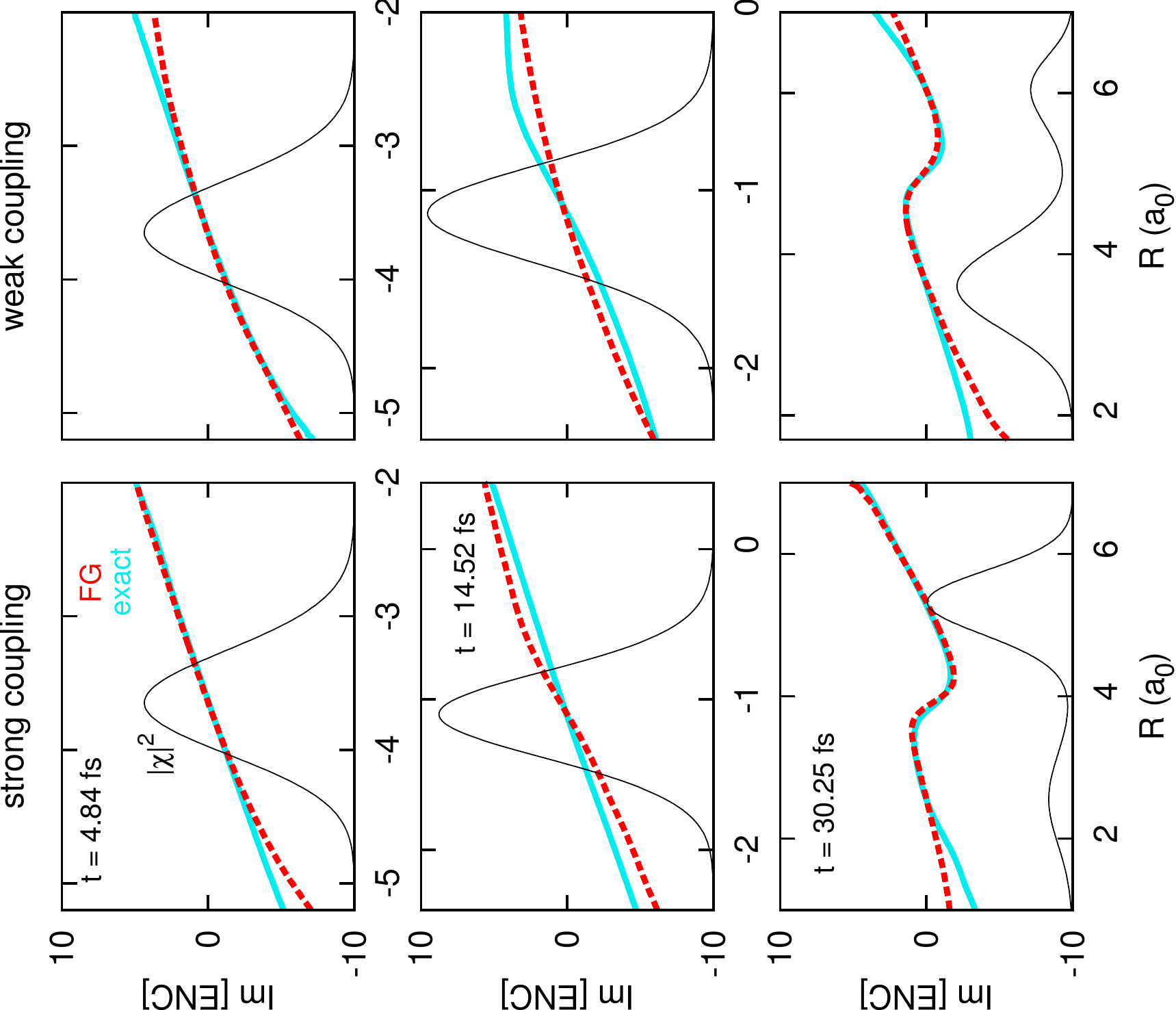}
\caption{Imaginary part of the ENC term for strong (left) and weak (right) non-adiabatic coupling strengths. The color code is the same as in Fig.~\ref{fig: Re ENC}.}
\label{fig: Im ENC}
\end{center}
\end{figure}
Fig.~\ref{fig: Im ENC} shows the imaginary part of the ENC term from Eq.~(\ref{eqn: polar nabla chi over chi}) at different time-steps during the dynamics, as in previous figures. The exact results (cyan) are compared only with the approximation (red) based on the semiclassical propagation of FGs. Even if we employ a simplified form of the Herman-Kluk propagator, where the pre-factor is set to 1, semiclassical results are in satisfactory good agreement with exact results.

One could consider improving the approach developed here by, for instance, explicitly computing the Herman-Kluk pre-factor, at the expenses of increasing the computational cost.

\section{Conclusion}\label{sec: conclusion}
We have reported our first semiclassical procedure adapted to the formalism of the exact factorization of the electron-nuclear wave function~\cite{AMG,AMG2}. The approach has been used to estimate the ENC term that explicitly depends on the nuclear wave function (gradient of its modulus and phase) in the electronic equation. In previous work~\cite{mqc,long_mqc} on the development of a mixed quantum-classical algorithm in the context of the exact factorization, such term has been treated fully classically and identified as the nuclear momentum. However, we observed that, despite the fact that this approximation is widely used in the literature~\cite{TSH_1990, truhlarFD2004, ciccottiJCP2000}, correction terms naturally arise in the factorization framework. The extent and role of the corrections may depend on the process that we intend to study, therefore we have shown in the present paper how to evaluate such corrections based on the semiclassical propagation of FGs, for a simple model case of non-adiabatic charge transfer. The main gain in using FGs, rather than a purely classical approach, lies in the possibility of obtaining smooth functions whose gradients can be easily determined without introducing large numerical errors.

The semiclassical results shown in the paper have been obtained by evolving FGs on the exact TDPES, that is known for the simple model system studied here as the outcome of exact calculations based on the numerical solution of the full TDSE. This procedure is a test, since the only approximation is the semiclassical treatment of the nuclear dynamics, whereas the electrons are treated exactly, via the information encoded in the TDPES. Moreover, this study provides a benchmark for future development, aiming at improving the initial, and lowest-order, mixed quantum-classical algorithm derived from the factorization~\cite{mqc,long_mqc}. The procedure described here can be easily implemented in an algorithm, as we will present elsewhere~\cite{MQC_min}, resulting in a novel mixed quantum-semiclassical scheme for solving coupled electron-nuclear dynamics.

\section*{Acknowledgements}
The authors would like to thank Neepa T. Maitra for her help in improving the presentation of the results. Partial support from the Deutsche Forschungsgemeinschaft (SFB 762) and from the European Commission (FP7-NMP-CRONOS) is gratefully acknowledged.


\begin{thebibliography}{74}%
\makeatletter
\providecommand \@ifxundefined [1]{%
 \@ifx{#1\undefined}
}%
\providecommand \@ifnum [1]{%
 \ifnum #1\expandafter \@firstoftwo
 \else \expandafter \@secondoftwo
 \fi
}%
\providecommand \@ifx [1]{%
 \ifx #1\expandafter \@firstoftwo
 \else \expandafter \@secondoftwo
 \fi
}%
\providecommand \natexlab [1]{#1}%
\providecommand \enquote  [1]{``#1''}%
\providecommand \bibnamefont  [1]{#1}%
\providecommand \bibfnamefont [1]{#1}%
\providecommand \citenamefont [1]{#1}%
\providecommand \href@noop [0]{\@secondoftwo}%
\providecommand \href [0]{\begingroup \@sanitize@url \@href}%
\providecommand \@href[1]{\@@startlink{#1}\@@href}%
\providecommand \@@href[1]{\endgroup#1\@@endlink}%
\providecommand \@sanitize@url [0]{\catcode `\\12\catcode `\$12\catcode
  `\&12\catcode `\#12\catcode `\^12\catcode `\_12\catcode `\%12\relax}%
\providecommand \@@startlink[1]{}%
\providecommand \@@endlink[0]{}%
\providecommand \url  [0]{\begingroup\@sanitize@url \@url }%
\providecommand \@url [1]{\endgroup\@href {#1}{\urlprefix }}%
\providecommand \urlprefix  [0]{URL }%
\providecommand \Eprint [0]{\href }%
\providecommand \doibase [0]{http://dx.doi.org/}%
\providecommand \selectlanguage [0]{\@gobble}%
\providecommand \bibinfo  [0]{\@secondoftwo}%
\providecommand \bibfield  [0]{\@secondoftwo}%
\providecommand \translation [1]{[#1]}%
\providecommand \BibitemOpen [0]{}%
\providecommand \bibitemStop [0]{}%
\providecommand \bibitemNoStop [0]{.\EOS\space}%
\providecommand \EOS [0]{\spacefactor3000\relax}%
\providecommand \BibitemShut  [1]{\csname bibitem#1\endcsname}%
\let\auto@bib@innerbib\@empty
\bibitem [{\citenamefont {Polli}\ \emph {et~al.}(2010)\citenamefont {Polli},
  \citenamefont {Alto\`e}, \citenamefont {Weingart}, \citenamefont {Spillane},
  \citenamefont {Manzoni}, \citenamefont {Brida}, \citenamefont {Tomasello},
  \citenamefont {Orlandi}, \citenamefont {Kukura}, \citenamefont {Mathies},
  \citenamefont {Garavelli},\ and\ \citenamefont {Cerullo}}]{cerulloN2010}%
  \BibitemOpen
  \bibfield  {author} {\bibinfo {author} {\bibfnamefont {D.}~\bibnamefont
  {Polli}}, \bibinfo {author} {\bibfnamefont {P.}~\bibnamefont {Alto\`e}},
  \bibinfo {author} {\bibfnamefont {O.}~\bibnamefont {Weingart}}, \bibinfo
  {author} {\bibfnamefont {K.~M.}\ \bibnamefont {Spillane}}, \bibinfo {author}
  {\bibfnamefont {C.}~\bibnamefont {Manzoni}}, \bibinfo {author} {\bibfnamefont
  {D.}~\bibnamefont {Brida}}, \bibinfo {author} {\bibfnamefont
  {G.}~\bibnamefont {Tomasello}}, \bibinfo {author} {\bibfnamefont
  {G.}~\bibnamefont {Orlandi}}, \bibinfo {author} {\bibfnamefont
  {P.}~\bibnamefont {Kukura}}, \bibinfo {author} {\bibfnamefont {R.~A.}\
  \bibnamefont {Mathies}}, \bibinfo {author} {\bibfnamefont {M.}~\bibnamefont
  {Garavelli}}, \ and\ \bibinfo {author} {\bibfnamefont {G.}~\bibnamefont
  {Cerullo}},\ }\href@noop {} {\bibfield  {journal} {\bibinfo  {journal}
  {Nature}\ }\textbf {\bibinfo {volume} {467}},\ \bibinfo {pages} {440}
  (\bibinfo {year} {2010})}\BibitemShut {NoStop}%
\bibitem [{\citenamefont {Chung}\ \emph {et~al.}(2012)\citenamefont {Chung},
  \citenamefont {Nanbu},\ and\ \citenamefont {Ishida}}]{ishidaJPCB2012}%
  \BibitemOpen
  \bibfield  {author} {\bibinfo {author} {\bibfnamefont {W.~C.}\ \bibnamefont
  {Chung}}, \bibinfo {author} {\bibfnamefont {S.}~\bibnamefont {Nanbu}}, \ and\
  \bibinfo {author} {\bibfnamefont {T.}~\bibnamefont {Ishida}},\ }\href@noop {}
  {\bibfield  {journal} {\bibinfo  {journal} {J. Phys. Chem. B}\ }\textbf
  {\bibinfo {volume} {116}},\ \bibinfo {pages} {8009} (\bibinfo {year}
  {2012})}\BibitemShut {NoStop}%
\bibitem [{\citenamefont {Rozzi}\ \emph {et~al.}(2013)\citenamefont {Rozzi},
  \citenamefont {Falke}, \citenamefont {Spallanzani}, \citenamefont {Rubio},
  \citenamefont {Molinari}, \citenamefont {Brida}, \citenamefont {Maiuri},
  \citenamefont {Cerullo}, \citenamefont {Schramm}, \citenamefont
  {Christoffers},\ and\ \citenamefont {Lienau}}]{rozziNC2013}%
  \BibitemOpen
  \bibfield  {author} {\bibinfo {author} {\bibfnamefont {C.~A.}\ \bibnamefont
  {Rozzi}}, \bibinfo {author} {\bibfnamefont {S.~M.}\ \bibnamefont {Falke}},
  \bibinfo {author} {\bibfnamefont {N.}~\bibnamefont {Spallanzani}}, \bibinfo
  {author} {\bibfnamefont {A.}~\bibnamefont {Rubio}}, \bibinfo {author}
  {\bibfnamefont {E.}~\bibnamefont {Molinari}}, \bibinfo {author}
  {\bibfnamefont {D.}~\bibnamefont {Brida}}, \bibinfo {author} {\bibfnamefont
  {M.}~\bibnamefont {Maiuri}}, \bibinfo {author} {\bibfnamefont
  {G.}~\bibnamefont {Cerullo}}, \bibinfo {author} {\bibfnamefont
  {H.}~\bibnamefont {Schramm}}, \bibinfo {author} {\bibfnamefont
  {J.}~\bibnamefont {Christoffers}}, \ and\ \bibinfo {author} {\bibfnamefont
  {C.}~\bibnamefont {Lienau}},\ }\href@noop {} {\bibfield  {journal} {\bibinfo
  {journal} {Nat. Communic.}\ }\textbf {\bibinfo {volume} {4}},\ \bibinfo
  {pages} {1602} (\bibinfo {year} {2013})}\BibitemShut {NoStop}%
\bibitem [{\citenamefont {Jailaubekov}\ \emph {et~al.}(2013)\citenamefont
  {Jailaubekov}, \citenamefont {Willard}, \citenamefont {Tritsch},
  \citenamefont {Chan}, \citenamefont {Sai}, \citenamefont {Gearba},
  \citenamefont {Kaake}, \citenamefont {Williams}, \citenamefont {Leung},
  \citenamefont {Rossky},\ and\ \citenamefont {Zhu}}]{jailaubekovNM2013}%
  \BibitemOpen
  \bibfield  {author} {\bibinfo {author} {\bibfnamefont {A.~E.}\ \bibnamefont
  {Jailaubekov}}, \bibinfo {author} {\bibfnamefont {A.~P.}\ \bibnamefont
  {Willard}}, \bibinfo {author} {\bibfnamefont {J.~R.}\ \bibnamefont
  {Tritsch}}, \bibinfo {author} {\bibfnamefont {W.-L.}\ \bibnamefont {Chan}},
  \bibinfo {author} {\bibfnamefont {N.}~\bibnamefont {Sai}}, \bibinfo {author}
  {\bibfnamefont {R.}~\bibnamefont {Gearba}}, \bibinfo {author} {\bibfnamefont
  {L.~G.}\ \bibnamefont {Kaake}}, \bibinfo {author} {\bibfnamefont {K.~J.}\
  \bibnamefont {Williams}}, \bibinfo {author} {\bibfnamefont {K.}~\bibnamefont
  {Leung}}, \bibinfo {author} {\bibfnamefont {P.~J.}\ \bibnamefont {Rossky}}, \
  and\ \bibinfo {author} {\bibfnamefont {X.-Y.}\ \bibnamefont {Zhu}},\
  }\href@noop {} {\bibfield  {journal} {\bibinfo  {journal} {Nat. Mater.}\
  }\textbf {\bibinfo {volume} {12}},\ \bibinfo {pages} {66} (\bibinfo {year}
  {2013})}\BibitemShut {NoStop}%
\bibitem [{\citenamefont {McEniry}\ \emph {et~al.}(2007)\citenamefont
  {McEniry}, \citenamefont {Bowler}, \citenamefont {Dundas}, \citenamefont
  {Horsfield}, \citenamefont {S{\'a}nchez},\ and\ \citenamefont
  {Todorov}}]{todorovJPCMN2007}%
  \BibitemOpen
  \bibfield  {author} {\bibinfo {author} {\bibfnamefont {E.~J.}\ \bibnamefont
  {McEniry}}, \bibinfo {author} {\bibfnamefont {D.~R.}\ \bibnamefont {Bowler}},
  \bibinfo {author} {\bibfnamefont {D.}~\bibnamefont {Dundas}}, \bibinfo
  {author} {\bibfnamefont {A.~P.}\ \bibnamefont {Horsfield}}, \bibinfo {author}
  {\bibfnamefont {C.~G.}\ \bibnamefont {S{\'a}nchez}}, \ and\ \bibinfo {author}
  {\bibfnamefont {T.~N.}\ \bibnamefont {Todorov}},\ }\href@noop {} {\bibfield
  {journal} {\bibinfo  {journal} {J. Phys.: Condens. Matter}\ }\textbf
  {\bibinfo {volume} {19}},\ \bibinfo {pages} {196201} (\bibinfo {year}
  {2007})}\BibitemShut {NoStop}%
\bibitem [{\citenamefont {Horsfield}\ \emph {et~al.}(2004)\citenamefont
  {Horsfield}, \citenamefont {Bowler}, \citenamefont {Fisher},\ and\
  \citenamefont {Todorov}}]{horsfield_JPhysCM2004}%
  \BibitemOpen
  \bibfield  {author} {\bibinfo {author} {\bibfnamefont {A.~P.}\ \bibnamefont
  {Horsfield}}, \bibinfo {author} {\bibfnamefont {D.~R.}\ \bibnamefont
  {Bowler}}, \bibinfo {author} {\bibfnamefont {A.~J.}\ \bibnamefont {Fisher}},
  \ and\ \bibinfo {author} {\bibfnamefont {T.~N.}\ \bibnamefont {Todorov}},\
  }\href@noop {} {\bibfield  {journal} {\bibinfo  {journal} {J. Phys.: Condens.
  Matter}\ }\textbf {\bibinfo {volume} {16}},\ \bibinfo {pages} {3609}
  (\bibinfo {year} {2004})}\BibitemShut {NoStop}%
\bibitem [{\citenamefont {Meyer}\ \emph {et~al.}(1990)\citenamefont {Meyer},
  \citenamefont {Manthe},\ and\ \citenamefont {Cederbaum}}]{cederbaumCPL1990}%
  \BibitemOpen
  \bibfield  {author} {\bibinfo {author} {\bibfnamefont {H.-D.}\ \bibnamefont
  {Meyer}}, \bibinfo {author} {\bibfnamefont {U.}~\bibnamefont {Manthe}}, \
  and\ \bibinfo {author} {\bibfnamefont {L.~S.}\ \bibnamefont {Cederbaum}},\
  }\href@noop {} {\bibfield  {journal} {\bibinfo  {journal} {Chem. Phys.
  Lett.}\ }\textbf {\bibinfo {volume} {165}},\ \bibinfo {pages} {73} (\bibinfo
  {year} {1990})}\BibitemShut {NoStop}%
\bibitem [{\citenamefont {Manthe}\ \emph {et~al.}(1992)\citenamefont {Manthe},
  \citenamefont {Meyer},\ and\ \citenamefont {Cederbaum}}]{cederbaumJCP1992}%
  \BibitemOpen
  \bibfield  {author} {\bibinfo {author} {\bibfnamefont {U.}~\bibnamefont
  {Manthe}}, \bibinfo {author} {\bibfnamefont {H.-D.}\ \bibnamefont {Meyer}}, \
  and\ \bibinfo {author} {\bibfnamefont {L.~S.}\ \bibnamefont {Cederbaum}},\
  }\href@noop {} {\bibfield  {journal} {\bibinfo  {journal} {J. Chem. Phys.}\
  }\textbf {\bibinfo {volume} {97}},\ \bibinfo {pages} {3199} (\bibinfo {year}
  {1992})}\BibitemShut {NoStop}%
\bibitem [{\citenamefont {Burghardt}\ \emph {et~al.}(1999)\citenamefont
  {Burghardt}, \citenamefont {Meyer},\ and\ \citenamefont
  {Cederbaum}}]{burghardtJCP1999}%
  \BibitemOpen
  \bibfield  {author} {\bibinfo {author} {\bibfnamefont {I.}~\bibnamefont
  {Burghardt}}, \bibinfo {author} {\bibfnamefont {H.-D.}\ \bibnamefont
  {Meyer}}, \ and\ \bibinfo {author} {\bibfnamefont {L.~S.}\ \bibnamefont
  {Cederbaum}},\ }\href@noop {} {\bibfield  {journal} {\bibinfo  {journal} {J.
  Chem. Phys.}\ }\textbf {\bibinfo {volume} {111}},\ \bibinfo {pages} {2927}
  (\bibinfo {year} {1999})}\BibitemShut {NoStop}%
\bibitem [{\citenamefont {Meyer}\ and\ \citenamefont
  {Worth}(2003)}]{worthTCA2003}%
  \BibitemOpen
  \bibfield  {author} {\bibinfo {author} {\bibfnamefont {H.-D.}\ \bibnamefont
  {Meyer}}\ and\ \bibinfo {author} {\bibfnamefont {G.~A.}\ \bibnamefont
  {Worth}},\ }\href@noop {} {\bibfield  {journal} {\bibinfo  {journal} {Theor.
  Chim. Acta}\ }\textbf {\bibinfo {volume} {109}},\ \bibinfo {pages} {251}
  (\bibinfo {year} {2003})}\BibitemShut {NoStop}%
\bibitem [{\citenamefont {Thoss}\ \emph {et~al.}(2004)\citenamefont {Thoss},
  \citenamefont {Domcke},\ and\ \citenamefont {Wang}}]{thossCM2004}%
  \BibitemOpen
  \bibfield  {author} {\bibinfo {author} {\bibfnamefont {M.}~\bibnamefont
  {Thoss}}, \bibinfo {author} {\bibfnamefont {W.}~\bibnamefont {Domcke}}, \
  and\ \bibinfo {author} {\bibfnamefont {H.}~\bibnamefont {Wang}},\ }\href@noop
  {} {\bibfield  {journal} {\bibinfo  {journal} {Chem. Phys.}\ }\textbf
  {\bibinfo {volume} {296}},\ \bibinfo {pages} {217 } (\bibinfo {year}
  {2004})}\BibitemShut {NoStop}%
\bibitem [{\citenamefont {L.~Wang}\ and\ \citenamefont
  {May}(2003)}]{mayJCP2006}%
  \BibitemOpen
  \bibfield  {author} {\bibinfo {author} {\bibfnamefont {H.-D.~M.}\
  \bibnamefont {L.~Wang}}\ and\ \bibinfo {author} {\bibfnamefont
  {V.}~\bibnamefont {May}},\ }\href@noop {} {\bibfield  {journal} {\bibinfo
  {journal} {J. Chem. Phys.}\ }\textbf {\bibinfo {volume} {125}},\ \bibinfo
  {pages} {014102} (\bibinfo {year} {2003})}\BibitemShut {NoStop}%
\bibitem [{\citenamefont {Westermann}\ \emph {et~al.}(2011)\citenamefont
  {Westermann}, \citenamefont {Brodbeck}, \citenamefont {Alexander B.
  Rozhenko~and},\ and\ \citenamefont {Manthe}}]{wastermannJCP2013}%
  \BibitemOpen
  \bibfield  {author} {\bibinfo {author} {\bibfnamefont {T.}~\bibnamefont
  {Westermann}}, \bibinfo {author} {\bibfnamefont {R.}~\bibnamefont
  {Brodbeck}}, \bibinfo {author} {\bibfnamefont {W.~S.}\ \bibnamefont
  {Alexander B. Rozhenko~and}}, \ and\ \bibinfo {author} {\bibfnamefont
  {U.}~\bibnamefont {Manthe}},\ }\href@noop {} {\bibfield  {journal} {\bibinfo
  {journal} {J. Chem. Phys.}\ }\textbf {\bibinfo {volume} {135}},\ \bibinfo
  {pages} {184102} (\bibinfo {year} {2011})}\BibitemShut {NoStop}%
\bibitem [{\citenamefont {Wang}\ and\ \citenamefont
  {Thoss}(2003)}]{thossJCP2003}%
  \BibitemOpen
  \bibfield  {author} {\bibinfo {author} {\bibfnamefont {H.}~\bibnamefont
  {Wang}}\ and\ \bibinfo {author} {\bibfnamefont {M.}~\bibnamefont {Thoss}},\
  }\href@noop {} {\bibfield  {journal} {\bibinfo  {journal} {J. Chem. Phys.}\
  }\textbf {\bibinfo {volume} {119}},\ \bibinfo {pages} {1289} (\bibinfo {year}
  {2003})}\BibitemShut {NoStop}%
\bibitem [{\citenamefont {Li}\ \emph {et~al.}(2010)\citenamefont {Li},
  \citenamefont {Kondov}, \citenamefont {Wang},\ and\ \citenamefont
  {Thoss}}]{thossJPCC2010}%
  \BibitemOpen
  \bibfield  {author} {\bibinfo {author} {\bibfnamefont {J.}~\bibnamefont
  {Li}}, \bibinfo {author} {\bibfnamefont {I.}~\bibnamefont {Kondov}}, \bibinfo
  {author} {\bibfnamefont {H.}~\bibnamefont {Wang}}, \ and\ \bibinfo {author}
  {\bibfnamefont {M.}~\bibnamefont {Thoss}},\ }\href@noop {} {\bibfield
  {journal} {\bibinfo  {journal} {J. Phys. Chem. C}\ }\textbf {\bibinfo
  {volume} {114}},\ \bibinfo {pages} {18481} (\bibinfo {year}
  {2010})}\BibitemShut {NoStop}%
\bibitem [{\citenamefont {Meng}\ \emph {et~al.}(2012)\citenamefont {Meng},
  \citenamefont {Faraji}, \citenamefont {Vendrell},\ and\ \citenamefont
  {Meyer}}]{mengJCP2012}%
  \BibitemOpen
  \bibfield  {author} {\bibinfo {author} {\bibfnamefont {Q.}~\bibnamefont
  {Meng}}, \bibinfo {author} {\bibfnamefont {S.}~\bibnamefont {Faraji}},
  \bibinfo {author} {\bibfnamefont {O.}~\bibnamefont {Vendrell}}, \ and\
  \bibinfo {author} {\bibfnamefont {H.-D.}\ \bibnamefont {Meyer}},\ }\href@noop
  {} {\bibfield  {journal} {\bibinfo  {journal} {J. Chem. Phys.}\ }\textbf
  {\bibinfo {volume} {137}},\ \bibinfo {pages} {134302} (\bibinfo {year}
  {2012})}\BibitemShut {NoStop}%
\bibitem [{\citenamefont {Schr{\"o}der}\ \emph {et~al.}(2008)\citenamefont
  {Schr{\"o}der}, \citenamefont {n~Macedo},\ and\ \citenamefont
  {Brown*}}]{brownPCCP}%
  \BibitemOpen
  \bibfield  {author} {\bibinfo {author} {\bibfnamefont {M.}~\bibnamefont
  {Schr{\"o}der}}, \bibinfo {author} {\bibfnamefont {J.-L.~C.}\ \bibnamefont
  {n~Macedo}}, \ and\ \bibinfo {author} {\bibfnamefont {A.}~\bibnamefont
  {Brown*}},\ }\href@noop {} {\bibfield  {journal} {\bibinfo  {journal} {Phys.
  Chem. Chem. Phys.}\ }\textbf {\bibinfo {volume} {10}},\ \bibinfo {pages}
  {850} (\bibinfo {year} {2008})}\BibitemShut {NoStop}%
\bibitem [{\citenamefont {Mart{\'i}nez}\ and\ \citenamefont
  {Levine}(1996)}]{martinezCPL1996}%
  \BibitemOpen
  \bibfield  {author} {\bibinfo {author} {\bibfnamefont {T.~J.}\ \bibnamefont
  {Mart{\'i}nez}}\ and\ \bibinfo {author} {\bibfnamefont {R.~D.}\ \bibnamefont
  {Levine}},\ }\href@noop {} {\bibfield  {journal} {\bibinfo  {journal} {Chem.
  Phys. Lett.}\ }\textbf {\bibinfo {volume} {259}},\ \bibinfo {pages} {252}
  (\bibinfo {year} {1996})}\BibitemShut {NoStop}%
\bibitem [{\citenamefont {Mart{\'i}nez}\ \emph {et~al.}(1996)\citenamefont
  {Mart{\'i}nez}, \citenamefont {Ben-Nun},\ and\ \citenamefont
  {Levine}}]{martinezJPC1996}%
  \BibitemOpen
  \bibfield  {author} {\bibinfo {author} {\bibfnamefont {T.~J.}\ \bibnamefont
  {Mart{\'i}nez}}, \bibinfo {author} {\bibfnamefont {M.}~\bibnamefont
  {Ben-Nun}}, \ and\ \bibinfo {author} {\bibfnamefont {R.~D.}\ \bibnamefont
  {Levine}},\ }\href@noop {} {\bibfield  {journal} {\bibinfo  {journal} {J.
  Phys. Chem.}\ }\textbf {\bibinfo {volume} {100}},\ \bibinfo {pages} {7884}
  (\bibinfo {year} {1996})}\BibitemShut {NoStop}%
\bibitem [{\citenamefont {Mart{\`i}nez}(2006)}]{martinezACR2006}%
  \BibitemOpen
  \bibfield  {author} {\bibinfo {author} {\bibfnamefont {T.~J.}\ \bibnamefont
  {Mart{\`i}nez}},\ }\href@noop {} {\bibfield  {journal} {\bibinfo  {journal}
  {Acc. Chem. Res.}\ }\textbf {\bibinfo {volume} {39}},\ \bibinfo {pages} {119}
  (\bibinfo {year} {2006})}\BibitemShut {NoStop}%
\bibitem [{\citenamefont {Ehrenfest}(1927)}]{ehrenfest}%
  \BibitemOpen
  \bibfield  {author} {\bibinfo {author} {\bibfnamefont {P.}~\bibnamefont
  {Ehrenfest}},\ }\href@noop {} {\bibfield  {journal} {\bibinfo  {journal}
  {Zeitschrift f{\"u}r Physik}\ }\textbf {\bibinfo {volume} {45}},\ \bibinfo
  {pages} {455} (\bibinfo {year} {1927})}\BibitemShut {NoStop}%
\bibitem [{\citenamefont {Pechukas}(1969)}]{pechukasPR1969_2}%
  \BibitemOpen
  \bibfield  {author} {\bibinfo {author} {\bibfnamefont {P.}~\bibnamefont
  {Pechukas}},\ }\href@noop {} {\bibfield  {journal} {\bibinfo  {journal}
  {Phys. Rev.}\ }\textbf {\bibinfo {volume} {181}},\ \bibinfo {pages} {174}
  (\bibinfo {year} {1969})}\BibitemShut {NoStop}%
\bibitem [{\citenamefont {Tully}(1990)}]{TSH_1990}%
  \BibitemOpen
  \bibfield  {author} {\bibinfo {author} {\bibfnamefont {J.~C.}\ \bibnamefont
  {Tully}},\ }\href@noop {} {\bibfield  {journal} {\bibinfo  {journal} {J.
  Chem. Phys.}\ }\textbf {\bibinfo {volume} {93}},\ \bibinfo {pages} {1061}
  (\bibinfo {year} {1990})}\BibitemShut {NoStop}%
\bibitem [{\citenamefont {Stock}\ and\ \citenamefont
  {Thoss}(1997)}]{Thoss_PRL1997}%
  \BibitemOpen
  \bibfield  {author} {\bibinfo {author} {\bibfnamefont {G.}~\bibnamefont
  {Stock}}\ and\ \bibinfo {author} {\bibfnamefont {M.}~\bibnamefont {Thoss}},\
  }\href@noop {} {\bibfield  {journal} {\bibinfo  {journal} {Phys. Rev. Lett.}\
  }\textbf {\bibinfo {volume} {78}},\ \bibinfo {pages} {578} (\bibinfo {year}
  {1997})}\BibitemShut {NoStop}%
\bibitem [{\citenamefont {Sun}\ and\ \citenamefont
  {Miller}(1997)}]{Miller_JCP1997_2}%
  \BibitemOpen
  \bibfield  {author} {\bibinfo {author} {\bibfnamefont {X.}~\bibnamefont
  {Sun}}\ and\ \bibinfo {author} {\bibfnamefont {W.~H.}\ \bibnamefont
  {Miller}},\ }\href@noop {} {\bibfield  {journal} {\bibinfo  {journal} {J.
  Chem. Phys.}\ }\textbf {\bibinfo {volume} {106}},\ \bibinfo {pages} {6346}
  (\bibinfo {year} {1997})}\BibitemShut {NoStop}%
\bibitem [{\citenamefont {Jasper}\ \emph {et~al.}(2004)\citenamefont {Jasper},
  \citenamefont {Zhu}, \citenamefont {Nangia},\ and\ \citenamefont
  {Truhlar}}]{truhlarFD2004}%
  \BibitemOpen
  \bibfield  {author} {\bibinfo {author} {\bibfnamefont {A.~W.}\ \bibnamefont
  {Jasper}}, \bibinfo {author} {\bibfnamefont {C.}~\bibnamefont {Zhu}},
  \bibinfo {author} {\bibfnamefont {S.}~\bibnamefont {Nangia}}, \ and\ \bibinfo
  {author} {\bibfnamefont {D.~G.}\ \bibnamefont {Truhlar}},\ }\href@noop {}
  {\bibfield  {journal} {\bibinfo  {journal} {Faraday Discuss.}\ }\textbf
  {\bibinfo {volume} {127}},\ \bibinfo {pages} {1} (\bibinfo {year}
  {2004})}\BibitemShut {NoStop}%
\bibitem [{\citenamefont {Nielsen}\ \emph {et~al.}(2000)\citenamefont
  {Nielsen}, \citenamefont {Kapral},\ and\ \citenamefont
  {Ciccotti}}]{ciccottiJCP2000}%
  \BibitemOpen
  \bibfield  {author} {\bibinfo {author} {\bibfnamefont {S.}~\bibnamefont
  {Nielsen}}, \bibinfo {author} {\bibfnamefont {R.}~\bibnamefont {Kapral}}, \
  and\ \bibinfo {author} {\bibfnamefont {G.}~\bibnamefont {Ciccotti}},\
  }\href@noop {} {\bibfield  {journal} {\bibinfo  {journal} {J. Chem. Phys.}\
  }\textbf {\bibinfo {volume} {112}},\ \bibinfo {pages} {6543} (\bibinfo {year}
  {2000})}\BibitemShut {NoStop}%
\bibitem [{\citenamefont {Hsieh}\ and\ \citenamefont
  {Kapral}(2013)}]{kapralJCP2013}%
  \BibitemOpen
  \bibfield  {author} {\bibinfo {author} {\bibfnamefont {C.-Y.}\ \bibnamefont
  {Hsieh}}\ and\ \bibinfo {author} {\bibfnamefont {R.}~\bibnamefont {Kapral}},\
  }\href@noop {} {\bibfield  {journal} {\bibinfo  {journal} {J. Chem. Phys.}\
  }\textbf {\bibinfo {volume} {138}},\ \bibinfo {pages} {134110} (\bibinfo
  {year} {2013})}\BibitemShut {NoStop}%
\bibitem [{\citenamefont {Bonella}\ and\ \citenamefont
  {Coker}(2005)}]{bonellaJCP2005}%
  \BibitemOpen
  \bibfield  {author} {\bibinfo {author} {\bibfnamefont {S.}~\bibnamefont
  {Bonella}}\ and\ \bibinfo {author} {\bibfnamefont {D.~F.}\ \bibnamefont
  {Coker}},\ }\href@noop {} {\bibfield  {journal} {\bibinfo  {journal} {J.
  Chem. Phys.}\ }\textbf {\bibinfo {volume} {122}},\ \bibinfo {pages} {194102}
  (\bibinfo {year} {2005})}\BibitemShut {NoStop}%
\bibitem [{\citenamefont {Huo}\ and\ \citenamefont
  {Coker}(2012)}]{cokerJCP2012}%
  \BibitemOpen
  \bibfield  {author} {\bibinfo {author} {\bibfnamefont {P.}~\bibnamefont
  {Huo}}\ and\ \bibinfo {author} {\bibfnamefont {D.~F.}\ \bibnamefont
  {Coker}},\ }\href@noop {} {\bibfield  {journal} {\bibinfo  {journal} {J.
  Chem. Phys.}\ }\textbf {\bibinfo {volume} {137}},\ \bibinfo {pages} {22A535}
  (\bibinfo {year} {2012})}\BibitemShut {NoStop}%
\bibitem [{\citenamefont {Doltsinis}\ and\ \citenamefont
  {Marx}(2002)}]{marxPRL2002}%
  \BibitemOpen
  \bibfield  {author} {\bibinfo {author} {\bibfnamefont {N.~L.}\ \bibnamefont
  {Doltsinis}}\ and\ \bibinfo {author} {\bibfnamefont {D.}~\bibnamefont
  {Marx}},\ }\href@noop {} {\bibfield  {journal} {\bibinfo  {journal} {Phys.
  Rev. lett.}\ }\textbf {\bibinfo {volume} {88}},\ \bibinfo {pages} {166402}
  (\bibinfo {year} {2002})}\BibitemShut {NoStop}%
\bibitem [{\citenamefont {Curchod}\ \emph {et~al.}(2011)\citenamefont
  {Curchod}, \citenamefont {Tavernelli},\ and\ \citenamefont
  {Rothlisberger}}]{tavernelliPCCP2011}%
  \BibitemOpen
  \bibfield  {author} {\bibinfo {author} {\bibfnamefont {B.~F.~E.}\
  \bibnamefont {Curchod}}, \bibinfo {author} {\bibfnamefont {I.}~\bibnamefont
  {Tavernelli}}, \ and\ \bibinfo {author} {\bibfnamefont {U.}~\bibnamefont
  {Rothlisberger}},\ }\href@noop {} {\bibfield  {journal} {\bibinfo  {journal}
  {Phys. Chem. Chem. Phys.}\ }\textbf {\bibinfo {volume} {13}},\ \bibinfo
  {pages} {3231} (\bibinfo {year} {2011})}\BibitemShut {NoStop}%
\bibitem [{\citenamefont {Wyatt}\ \emph {et~al.}(2001)\citenamefont {Wyatt},
  \citenamefont {Lopreore},\ and\ \citenamefont {Parlant}}]{wyattJCP2001}%
  \BibitemOpen
  \bibfield  {author} {\bibinfo {author} {\bibfnamefont {R.~E.}\ \bibnamefont
  {Wyatt}}, \bibinfo {author} {\bibfnamefont {C.~L.}\ \bibnamefont {Lopreore}},
  \ and\ \bibinfo {author} {\bibfnamefont {G.}~\bibnamefont {Parlant}},\
  }\href@noop {} {\bibfield  {journal} {\bibinfo  {journal} {J. Chem. Phys.}\
  }\textbf {\bibinfo {volume} {114}},\ \bibinfo {pages} {5113} (\bibinfo {year}
  {2001})}\BibitemShut {NoStop}%
\bibitem [{\citenamefont {Burghardt}(2005)}]{burghardtJCP2005}%
  \BibitemOpen
  \bibfield  {author} {\bibinfo {author} {\bibfnamefont {I.}~\bibnamefont
  {Burghardt}},\ }\href@noop {} {\bibfield  {journal} {\bibinfo  {journal} {J.
  Chem. Phys.}\ }\textbf {\bibinfo {volume} {122}},\ \bibinfo {pages} {094103}
  (\bibinfo {year} {2005})}\BibitemShut {NoStop}%
\bibitem [{\citenamefont {Prezhdo}\ and\ \citenamefont
  {Brooksby}(2001)}]{prezhdoPRL2001}%
  \BibitemOpen
  \bibfield  {author} {\bibinfo {author} {\bibfnamefont {O.~V.}\ \bibnamefont
  {Prezhdo}}\ and\ \bibinfo {author} {\bibfnamefont {C.}~\bibnamefont
  {Brooksby}},\ }\href@noop {} {\bibfield  {journal} {\bibinfo  {journal}
  {Phys. Rev. Lett.}\ }\textbf {\bibinfo {volume} {86}},\ \bibinfo {pages}
  {3215} (\bibinfo {year} {2001})}\BibitemShut {NoStop}%
\bibitem [{\citenamefont {Zamstein}\ and\ \citenamefont
  {Tannor}(2012)}]{tannorJCP2012}%
  \BibitemOpen
  \bibfield  {author} {\bibinfo {author} {\bibfnamefont {N.}~\bibnamefont
  {Zamstein}}\ and\ \bibinfo {author} {\bibfnamefont {D.~J.}\ \bibnamefont
  {Tannor}},\ }\href@noop {} {\bibfield  {journal} {\bibinfo  {journal} {J.
  Chem. Phys.}\ }\textbf {\bibinfo {volume} {137}},\ \bibinfo {pages} {22A518}
  (\bibinfo {year} {2012})}\BibitemShut {NoStop}%
\bibitem [{\citenamefont {Bousquet}\ \emph {et~al.}(2011)\citenamefont
  {Bousquet}, \citenamefont {Hughes}, \citenamefont {Micha},\ and\
  \citenamefont {Burghardt}}]{burghardtJCP2011}%
  \BibitemOpen
  \bibfield  {author} {\bibinfo {author} {\bibfnamefont {D.}~\bibnamefont
  {Bousquet}}, \bibinfo {author} {\bibfnamefont {K.~H.}\ \bibnamefont
  {Hughes}}, \bibinfo {author} {\bibfnamefont {D.~A.}\ \bibnamefont {Micha}}, \
  and\ \bibinfo {author} {\bibfnamefont {I.}~\bibnamefont {Burghardt}},\
  }\href@noop {} {\bibfield  {journal} {\bibinfo  {journal} {J. Chem. Phys.}\
  }\textbf {\bibinfo {volume} {134}},\ \bibinfo {pages} {064116} (\bibinfo
  {year} {2011})}\BibitemShut {NoStop}%
\bibitem [{\citenamefont {Richardson}\ and\ \citenamefont
  {Thoss}(2013)}]{thossJCP2013}%
  \BibitemOpen
  \bibfield  {author} {\bibinfo {author} {\bibfnamefont {J.~O.}\ \bibnamefont
  {Richardson}}\ and\ \bibinfo {author} {\bibfnamefont {M.}~\bibnamefont
  {Thoss}},\ }\href@noop {} {\bibfield  {journal} {\bibinfo  {journal} {J.
  Chem. Phys.}\ }\textbf {\bibinfo {volume} {139}},\ \bibinfo {pages} {031102}
  (\bibinfo {year} {2013})}\BibitemShut {NoStop}%
\bibitem [{\citenamefont {Ananth}(2013)}]{ananthJCP2013}%
  \BibitemOpen
  \bibfield  {author} {\bibinfo {author} {\bibfnamefont {N.}~\bibnamefont
  {Ananth}},\ }\href@noop {} {\bibfield  {journal} {\bibinfo  {journal} {J.
  Chem. Phys.}\ }\textbf {\bibinfo {volume} {139}},\ \bibinfo {pages} {124102}
  (\bibinfo {year} {2013})}\BibitemShut {NoStop}%
\bibitem [{\citenamefont {Abedi}\ \emph {et~al.}(2010)\citenamefont {Abedi},
  \citenamefont {Maitra},\ and\ \citenamefont {Gross}}]{AMG}%
  \BibitemOpen
  \bibfield  {author} {\bibinfo {author} {\bibfnamefont {A.}~\bibnamefont
  {Abedi}}, \bibinfo {author} {\bibfnamefont {N.~T.}\ \bibnamefont {Maitra}}, \
  and\ \bibinfo {author} {\bibfnamefont {E.~K.~U.}\ \bibnamefont {Gross}},\
  }\href@noop {} {\bibfield  {journal} {\bibinfo  {journal} {Phys. Rev. Lett.}\
  }\textbf {\bibinfo {volume} {105}},\ \bibinfo {pages} {123002} (\bibinfo
  {year} {2010})}\BibitemShut {NoStop}%
\bibitem [{\citenamefont {Abedi}\ \emph {et~al.}(2012)\citenamefont {Abedi},
  \citenamefont {Maitra},\ and\ \citenamefont {Gross}}]{AMG2}%
  \BibitemOpen
  \bibfield  {author} {\bibinfo {author} {\bibfnamefont {A.}~\bibnamefont
  {Abedi}}, \bibinfo {author} {\bibfnamefont {N.~T.}\ \bibnamefont {Maitra}}, \
  and\ \bibinfo {author} {\bibfnamefont {E.~K.~U.}\ \bibnamefont {Gross}},\
  }\href@noop {} {\bibfield  {journal} {\bibinfo  {journal} {J. Chem. Phys.}\
  }\textbf {\bibinfo {volume} {137}},\ \bibinfo {pages} {22A530} (\bibinfo
  {year} {2012})}\BibitemShut {NoStop}%
\bibitem [{\citenamefont {Abedi}\ \emph
  {et~al.}(2013{\natexlab{a}})\citenamefont {Abedi}, \citenamefont {Agostini},
  \citenamefont {Suzuki},\ and\ \citenamefont {Gross}}]{steps}%
  \BibitemOpen
  \bibfield  {author} {\bibinfo {author} {\bibfnamefont {A.}~\bibnamefont
  {Abedi}}, \bibinfo {author} {\bibfnamefont {F.}~\bibnamefont {Agostini}},
  \bibinfo {author} {\bibfnamefont {Y.}~\bibnamefont {Suzuki}}, \ and\ \bibinfo
  {author} {\bibfnamefont {E.~K.~U.}\ \bibnamefont {Gross}},\ }\href@noop {}
  {\bibfield  {journal} {\bibinfo  {journal} {Phys. Rev. Lett}\ }\textbf
  {\bibinfo {volume} {110}},\ \bibinfo {pages} {263001} (\bibinfo {year}
  {2013}{\natexlab{a}})}\BibitemShut {NoStop}%
\bibitem [{\citenamefont {Agostini}\ \emph {et~al.}(2013)\citenamefont
  {Agostini}, \citenamefont {Abedi}, \citenamefont {Suzuki},\ and\
  \citenamefont {Gross}}]{long_steps}%
  \BibitemOpen
  \bibfield  {author} {\bibinfo {author} {\bibfnamefont {F.}~\bibnamefont
  {Agostini}}, \bibinfo {author} {\bibfnamefont {A.}~\bibnamefont {Abedi}},
  \bibinfo {author} {\bibfnamefont {Y.}~\bibnamefont {Suzuki}}, \ and\ \bibinfo
  {author} {\bibfnamefont {E.~K.~U.}\ \bibnamefont {Gross}},\ }\href@noop {}
  {\bibfield  {journal} {\bibinfo  {journal} {Mol. Phys.}\ }\textbf {\bibinfo
  {volume} {111}},\ \bibinfo {pages} {3625} (\bibinfo {year}
  {2013})}\BibitemShut {NoStop}%
\bibitem [{\citenamefont {Agostini}\ \emph
  {et~al.}(2014{\natexlab{a}})\citenamefont {Agostini}, \citenamefont {Abedi},
  \citenamefont {Suzuki}, \citenamefont {Min}, \citenamefont {Maitra},\ and\
  \citenamefont {Gross}}]{long_steps_mt}%
  \BibitemOpen
  \bibfield  {author} {\bibinfo {author} {\bibfnamefont {F.}~\bibnamefont
  {Agostini}}, \bibinfo {author} {\bibfnamefont {A.}~\bibnamefont {Abedi}},
  \bibinfo {author} {\bibfnamefont {Y.}~\bibnamefont {Suzuki}}, \bibinfo
  {author} {\bibfnamefont {S.~K.}\ \bibnamefont {Min}}, \bibinfo {author}
  {\bibfnamefont {N.~T.}\ \bibnamefont {Maitra}}, \ and\ \bibinfo {author}
  {\bibfnamefont {E.~K.~U.}\ \bibnamefont {Gross}},\ }\href@noop {} {\bibfield
  {journal} {\bibinfo  {journal} {arXiv:1406.4667 [physics.chem-ph]}\ }
  (\bibinfo {year} {2014}{\natexlab{a}})}\BibitemShut {NoStop}%
\bibitem [{\citenamefont {Abedi}\ \emph {et~al.}(2014)\citenamefont {Abedi},
  \citenamefont {Agostini},\ and\ \citenamefont {Gross}}]{mqc}%
  \BibitemOpen
  \bibfield  {author} {\bibinfo {author} {\bibfnamefont {A.}~\bibnamefont
  {Abedi}}, \bibinfo {author} {\bibfnamefont {F.}~\bibnamefont {Agostini}}, \
  and\ \bibinfo {author} {\bibfnamefont {E.~K.~U.}\ \bibnamefont {Gross}},\
  }\href@noop {} {\bibfield  {journal} {\bibinfo  {journal} {Europhys. Lett.}\
  }\textbf {\bibinfo {volume} {106}},\ \bibinfo {pages} {33001} (\bibinfo
  {year} {2014})}\BibitemShut {NoStop}%
\bibitem [{\citenamefont {Agostini}\ \emph
  {et~al.}(2014{\natexlab{b}})\citenamefont {Agostini}, \citenamefont {Abedi},\
  and\ \citenamefont {Gross}}]{long_mqc}%
  \BibitemOpen
  \bibfield  {author} {\bibinfo {author} {\bibfnamefont {F.}~\bibnamefont
  {Agostini}}, \bibinfo {author} {\bibfnamefont {A.}~\bibnamefont {Abedi}}, \
  and\ \bibinfo {author} {\bibfnamefont {E.~K.~U.}\ \bibnamefont {Gross}},\
  }\href@noop {} {\bibfield  {journal} {\bibinfo  {journal} {J. Chem. Phys.}\
  }\textbf {\bibinfo {volume} {141}},\ \bibinfo {pages} {214101} (\bibinfo
  {year} {2014}{\natexlab{b}})}\BibitemShut {NoStop}%
\bibitem [{\citenamefont {Vleck}(1928)}]{VanVleck_PNAS1928}%
  \BibitemOpen
  \bibfield  {author} {\bibinfo {author} {\bibfnamefont {J.~H.~V.}\
  \bibnamefont {Vleck}},\ }\href@noop {} {\bibfield  {journal} {\bibinfo
  {journal} {Proc. Nat. Ac. Sci.}\ }\textbf {\bibinfo {volume} {14}},\ \bibinfo
  {pages} {178} (\bibinfo {year} {1928})}\BibitemShut {NoStop}%
\bibitem [{\citenamefont {Min}\ \emph {et~al.}(tted)\citenamefont {Min},
  \citenamefont {Agostini},\ and\ \citenamefont {Gross}}]{MQC_min}%
  \BibitemOpen
  \bibfield  {author} {\bibinfo {author} {\bibfnamefont {S.~K.}\ \bibnamefont
  {Min}}, \bibinfo {author} {\bibfnamefont {F.}~\bibnamefont {Agostini}}, \
  and\ \bibinfo {author} {\bibfnamefont {E.~K.~U.}\ \bibnamefont {Gross}},\
  }\href@noop {} {\  (\bibinfo {year} {to be submitted})}\BibitemShut {NoStop}%
\bibitem [{\citenamefont {Heller}(1981)}]{Heller_JCP1981_2}%
  \BibitemOpen
  \bibfield  {author} {\bibinfo {author} {\bibfnamefont {E.~J.}\ \bibnamefont
  {Heller}},\ }\href@noop {} {\bibfield  {journal} {\bibinfo  {journal} {J.
  Chem. Phys.}\ }\textbf {\bibinfo {volume} {75}},\ \bibinfo {pages} {2923}
  (\bibinfo {year} {1981})}\BibitemShut {NoStop}%
\bibitem [{\citenamefont {Herman}\ and\ \citenamefont
  {Kluk}(1984)}]{Kluk_CP1984}%
  \BibitemOpen
  \bibfield  {author} {\bibinfo {author} {\bibfnamefont {M.~F.}\ \bibnamefont
  {Herman}}\ and\ \bibinfo {author} {\bibfnamefont {E.}~\bibnamefont {Kluk}},\
  }\href@noop {} {\bibfield  {journal} {\bibinfo  {journal} {Chem. Phys.}\
  }\textbf {\bibinfo {volume} {91}},\ \bibinfo {pages} {27} (\bibinfo {year}
  {1984})}\BibitemShut {NoStop}%
\bibitem [{\citenamefont {Kluk}\ \emph {et~al.}(1986)\citenamefont {Kluk},
  \citenamefont {Herman},\ and\ \citenamefont {Davis}}]{Davis_JCP1986}%
  \BibitemOpen
  \bibfield  {author} {\bibinfo {author} {\bibfnamefont {E.}~\bibnamefont
  {Kluk}}, \bibinfo {author} {\bibfnamefont {M.~F.}\ \bibnamefont {Herman}}, \
  and\ \bibinfo {author} {\bibfnamefont {H.~L.}\ \bibnamefont {Davis}},\
  }\href@noop {} {\bibfield  {journal} {\bibinfo  {journal} {J. Chem. Phys.}\
  }\textbf {\bibinfo {volume} {84}},\ \bibinfo {pages} {326} (\bibinfo {year}
  {1986})}\BibitemShut {NoStop}%
\bibitem [{\citenamefont {Kay}(2006)}]{Kay_CP2006}%
  \BibitemOpen
  \bibfield  {author} {\bibinfo {author} {\bibfnamefont {K.~G.}\ \bibnamefont
  {Kay}},\ }\href@noop {} {\bibfield  {journal} {\bibinfo  {journal} {Chem.
  Phys.}\ }\textbf {\bibinfo {volume} {322}},\ \bibinfo {pages} {3} (\bibinfo
  {year} {2006})}\BibitemShut {NoStop}%
\bibitem [{\citenamefont {Miller}(2002)}]{Miller_MP2002}%
  \BibitemOpen
  \bibfield  {author} {\bibinfo {author} {\bibfnamefont {W.~H.}\ \bibnamefont
  {Miller}},\ }\href@noop {} {\bibfield  {journal} {\bibinfo  {journal} {Mol.
  Phys.}\ }\textbf {\bibinfo {volume} {100}},\ \bibinfo {pages} {397} (\bibinfo
  {year} {2002})}\BibitemShut {NoStop}%
\bibitem [{\citenamefont {Miller}(1970)}]{Miller_JCP1970}%
  \BibitemOpen
  \bibfield  {author} {\bibinfo {author} {\bibfnamefont {W.~H.}\ \bibnamefont
  {Miller}},\ }\href@noop {} {\bibfield  {journal} {\bibinfo  {journal} {J.
  Chem. Phys.}\ }\textbf {\bibinfo {volume} {53}},\ \bibinfo {pages} {3578}
  (\bibinfo {year} {1970})}\BibitemShut {NoStop}%
\bibitem [{\citenamefont {Thoss}\ and\ \citenamefont
  {Wang}(2004)}]{Wang_ARPC2004}%
  \BibitemOpen
  \bibfield  {author} {\bibinfo {author} {\bibfnamefont {M.}~\bibnamefont
  {Thoss}}\ and\ \bibinfo {author} {\bibfnamefont {H.}~\bibnamefont {Wang}},\
  }\href@noop {} {\bibfield  {journal} {\bibinfo  {journal} {Ann. Rev. Phys.
  Chem.}\ }\textbf {\bibinfo {volume} {55}},\ \bibinfo {pages} {299} (\bibinfo
  {year} {2004})}\BibitemShut {NoStop}%
\bibitem [{\citenamefont {Kay}(2005)}]{Kay_ARPC2005}%
  \BibitemOpen
  \bibfield  {author} {\bibinfo {author} {\bibfnamefont {K.~G.}\ \bibnamefont
  {Kay}},\ }\href@noop {} {\bibfield  {journal} {\bibinfo  {journal} {Ann. Rev.
  Phys. Chem.}\ }\textbf {\bibinfo {volume} {56}},\ \bibinfo {pages} {255}
  (\bibinfo {year} {2005})}\BibitemShut {NoStop}%
\bibitem [{\citenamefont {Miller}(2001)}]{Miller_JPCA2001}%
  \BibitemOpen
  \bibfield  {author} {\bibinfo {author} {\bibfnamefont {W.~H.}\ \bibnamefont
  {Miller}},\ }\href@noop {} {\bibfield  {journal} {\bibinfo  {journal} {J.
  Phys. Chem A}\ }\textbf {\bibinfo {volume} {105}},\ \bibinfo {pages} {2942}
  (\bibinfo {year} {2001})}\BibitemShut {NoStop}%
\bibitem [{\citenamefont {Shin}\ and\ \citenamefont {Metiu}(1995)}]{MM}%
  \BibitemOpen
  \bibfield  {author} {\bibinfo {author} {\bibfnamefont {S.}~\bibnamefont
  {Shin}}\ and\ \bibinfo {author} {\bibfnamefont {H.}~\bibnamefont {Metiu}},\
  }\href@noop {} {\bibfield  {journal} {\bibinfo  {journal} {J. Chem. Phys.}\
  }\textbf {\bibinfo {volume} {102}},\ \bibinfo {pages} {23} (\bibinfo {year}
  {1995})}\BibitemShut {NoStop}%
\bibitem [{\citenamefont {Hunter}(1975)}]{hunter}%
  \BibitemOpen
  \bibfield  {author} {\bibinfo {author} {\bibfnamefont {G.}~\bibnamefont
  {Hunter}},\ }\href@noop {} {\bibfield  {journal} {\bibinfo  {journal} {Int.
  J. Quantum Chem}\ }\textbf {\bibinfo {volume} {9}},\ \bibinfo {pages} {237}
  (\bibinfo {year} {1975})}\BibitemShut {NoStop}%
\bibitem [{\citenamefont {Gidopoulos}\ and\ \citenamefont
  {Gross}(2014)}]{Gross_PTRSA2014}%
  \BibitemOpen
  \bibfield  {author} {\bibinfo {author} {\bibfnamefont {N.~I.}\ \bibnamefont
  {Gidopoulos}}\ and\ \bibinfo {author} {\bibfnamefont {E.~K.~U.}\ \bibnamefont
  {Gross}},\ }\href@noop {} {\bibfield  {journal} {\bibinfo  {journal} {Phil.
  Trans. R. Soc. A}\ }\textbf {\bibinfo {volume} {372}},\ \bibinfo {pages}
  {20130059} (\bibinfo {year} {2014})}\BibitemShut {NoStop}%
\bibitem [{\citenamefont {Frenkel}(1934)}]{frenkel}%
  \BibitemOpen
  \bibfield  {author} {\bibinfo {author} {\bibfnamefont {J.}~\bibnamefont
  {Frenkel}},\ }\href@noop {} {\emph {\bibinfo {title} {Wave mechanics}}},\
  \bibinfo {edition} {{C}larendon, {O}xford}\ ed.\ (\bibinfo {year}
  {1934})\BibitemShut {NoStop}%
\bibitem [{\citenamefont {McLachlan}(1964)}]{mclachlan}%
  \BibitemOpen
  \bibfield  {author} {\bibinfo {author} {\bibfnamefont {A.~D.}\ \bibnamefont
  {McLachlan}},\ }\href@noop {} {\bibfield  {journal} {\bibinfo  {journal}
  {Mol. Phys.}\ }\textbf {\bibinfo {volume} {8}},\ \bibinfo {pages} {39}
  (\bibinfo {year} {1964})}\BibitemShut {NoStop}%
\bibitem [{\citenamefont {Alonso}\ \emph {et~al.}(2013)\citenamefont {Alonso},
  \citenamefont {Clemente-Gallardo}, \citenamefont {Echeniche-Robba},\ and\
  \citenamefont {Jover-Galtier}}]{alonsoJCP2013}%
  \BibitemOpen
  \bibfield  {author} {\bibinfo {author} {\bibfnamefont {J.~L.}\ \bibnamefont
  {Alonso}}, \bibinfo {author} {\bibfnamefont {J.}~\bibnamefont
  {Clemente-Gallardo}}, \bibinfo {author} {\bibfnamefont {P.}~\bibnamefont
  {Echeniche-Robba}}, \ and\ \bibinfo {author} {\bibfnamefont {J.~A.}\
  \bibnamefont {Jover-Galtier}},\ }\href@noop {} {\bibfield  {journal}
  {\bibinfo  {journal} {J. Chem. Phys.}\ }\textbf {\bibinfo {volume} {139}},\
  \bibinfo {pages} {087101} (\bibinfo {year} {2013})}\BibitemShut {NoStop}%
\bibitem [{\citenamefont {Abedi}\ \emph
  {et~al.}(2013{\natexlab{b}})\citenamefont {Abedi}, \citenamefont {Maitra},\
  and\ \citenamefont {Gross}}]{AMG2013}%
  \BibitemOpen
  \bibfield  {author} {\bibinfo {author} {\bibfnamefont {A.}~\bibnamefont
  {Abedi}}, \bibinfo {author} {\bibfnamefont {N.~T.}\ \bibnamefont {Maitra}}, \
  and\ \bibinfo {author} {\bibfnamefont {E.~K.~U.}\ \bibnamefont {Gross}},\
  }\href@noop {} {\bibfield  {journal} {\bibinfo  {journal} {J. Chem. Phys.}\
  }\textbf {\bibinfo {volume} {139}},\ \bibinfo {pages} {087102} (\bibinfo
  {year} {2013}{\natexlab{b}})}\BibitemShut {NoStop}%
\bibitem [{\citenamefont {Ghosh}\ and\ \citenamefont
  {Dhara}(1988)}]{Ghosh-Dhara}%
  \BibitemOpen
  \bibfield  {author} {\bibinfo {author} {\bibfnamefont {S.~K.}\ \bibnamefont
  {Ghosh}}\ and\ \bibinfo {author} {\bibfnamefont {A.~K.}\ \bibnamefont
  {Dhara}},\ }\href@noop {} {\bibfield  {journal} {\bibinfo  {journal} {Phys.
  Rev. A}\ }\textbf {\bibinfo {volume} {38}},\ \bibinfo {pages} {1149}
  (\bibinfo {year} {1988})}\BibitemShut {NoStop}%
\bibitem [{\citenamefont {Runge}\ and\ \citenamefont {Gross}(1984)}]{RGT}%
  \BibitemOpen
  \bibfield  {author} {\bibinfo {author} {\bibfnamefont {E.}~\bibnamefont
  {Runge}}\ and\ \bibinfo {author} {\bibfnamefont {E.~K.~U.}\ \bibnamefont
  {Gross}},\ }\href@noop {} {\bibfield  {journal} {\bibinfo  {journal} {Phys.
  Rev. Lett.}\ }\textbf {\bibinfo {volume} {52}},\ \bibinfo {pages} {997}
  (\bibinfo {year} {1984})}\BibitemShut {NoStop}%
\bibitem [{\citenamefont {Min}\ \emph {et~al.}(2014)\citenamefont {Min},
  \citenamefont {Abedi}, \citenamefont {Kim},\ and\ \citenamefont
  {Gross}}]{CI_MAG}%
  \BibitemOpen
  \bibfield  {author} {\bibinfo {author} {\bibfnamefont {S.~K.}\ \bibnamefont
  {Min}}, \bibinfo {author} {\bibfnamefont {A.}~\bibnamefont {Abedi}}, \bibinfo
  {author} {\bibfnamefont {K.~S.}\ \bibnamefont {Kim}}, \ and\ \bibinfo
  {author} {\bibfnamefont {E.~K.~U.}\ \bibnamefont {Gross}},\ }\href@noop {}
  {\bibfield  {journal} {\bibinfo  {journal} {Phys. Rev. Lett.}\ }\textbf
  {\bibinfo {volume} {113}},\ \bibinfo {pages} {263004} (\bibinfo {year}
  {2014})}\BibitemShut {NoStop}%
\bibitem [{Note1()}]{Note1}%
  \BibitemOpen
  \bibinfo {note} {To be precise, the nuclear density is approximated, within
  the classical treatment, as $\delta $-function, centered at all times at the
  classical trajectory. But this contribution is totally omitted in the
  expression of the ENC term.}\BibitemShut {Stop}%
\bibitem [{\citenamefont {Feit}\ \emph {et~al.}(1982)\citenamefont {Feit},
  \citenamefont {{Fleck~Jr.}},\ and\ \citenamefont {Steiger}}]{spo}%
  \BibitemOpen
  \bibfield  {author} {\bibinfo {author} {\bibfnamefont {M.~D.}\ \bibnamefont
  {Feit}}, \bibinfo {author} {\bibfnamefont {F.~A.}\ \bibnamefont
  {{Fleck~Jr.}}}, \ and\ \bibinfo {author} {\bibfnamefont {A.}~\bibnamefont
  {Steiger}},\ }\href@noop {} {\bibfield  {journal} {\bibinfo  {journal} {J.
  Comput. Phys.}\ }\textbf {\bibinfo {volume} {47}},\ \bibinfo {pages} {412}
  (\bibinfo {year} {1982})}\BibitemShut {NoStop}%
\bibitem [{Note2()}]{Note2}%
  \BibitemOpen
  \bibinfo {note} {The value of $\gamma $ is chosen to have the best overlap
  between the initial semiclassical density and the exact density, in order to
  impose initial conditions as close as possible to the exact
  ones.}\BibitemShut {Stop}%
\bibitem [{\citenamefont {Gelman}\ and\ \citenamefont
  {Schwartz}(2008)}]{:/content/aip/journal/jcp/129/2/10.1063/1.2949818}%
  \BibitemOpen
  \bibfield  {author} {\bibinfo {author} {\bibfnamefont {D.}~\bibnamefont
  {Gelman}}\ and\ \bibinfo {author} {\bibfnamefont {S.~D.}\ \bibnamefont
  {Schwartz}},\ }\href@noop {} {\bibfield  {journal} {\bibinfo  {journal} {J.
  Chem. Phys.}\ }\textbf {\bibinfo {volume} {129}},\ \bibinfo {pages} {024504}
  (\bibinfo {year} {2008})}\BibitemShut {NoStop}%
\bibitem [{\citenamefont {Ankerhold}(2007)}]{ankerhold_book}%
  \BibitemOpen
  \bibfield  {author} {\bibinfo {author} {\bibfnamefont {J.}~\bibnamefont
  {Ankerhold}},\ }\href@noop {} {\emph {\bibinfo {title} {Quantum tunnelling in
  complex systems. {T}he semiclassical approach}}},\ \bibinfo {series}
  {{S}pringer {T}racts in {M}odern {P}hysics}, Vol.\ \bibinfo {volume} {224}\
  (\bibinfo  {publisher} {Springer-Verlag Berlin Heidelberg},\ \bibinfo {year}
  {2007})\BibitemShut {NoStop}%
\bibitem [{\citenamefont {Saltzer}\ and\ \citenamefont
  {Ankerhold}(2003)}]{AnkerholdPRA2003}%
  \BibitemOpen
  \bibfield  {author} {\bibinfo {author} {\bibfnamefont {M.}~\bibnamefont
  {Saltzer}}\ and\ \bibinfo {author} {\bibfnamefont {J.}~\bibnamefont
  {Ankerhold}},\ }\href@noop {} {\bibfield  {journal} {\bibinfo  {journal}
  {Phys. Rev. A}\ }\textbf {\bibinfo {volume} {68}},\ \bibinfo {pages} {042108}
  (\bibinfo {year} {2003})}\BibitemShut {NoStop}%
\bibitem [{\citenamefont {Kay}(2013)}]{KayPRA2013}%
  \BibitemOpen
  \bibfield  {author} {\bibinfo {author} {\bibfnamefont {K.~G.}\ \bibnamefont
  {Kay}},\ }\href@noop {} {\bibfield  {journal} {\bibinfo  {journal} {Phys.
  Rev. A}\ }\textbf {\bibinfo {volume} {88}},\ \bibinfo {pages} {012122}
  (\bibinfo {year} {2013})}\BibitemShut {NoStop}%
\end{thebibliography}
%

\end{document}